\documentstyle[12pt,epsfig,color]{article}
\def\baselinestretch{1.3}
\newcommand{\comment}[1]{}

\def\beq{\begin{equation}}
\def\eeq{\end{equation}}
\def\beqn{\begin{eqnarray}}
\def\eeqn{\end{eqnarray}}

\def\mygraph#1#2{ \subfigure[]{
\label{#1}
\hspace*{-0.6in}
\centering
\hspace*{4ex}
\includegraphics[width=0.9\textwidth,height=0.7\textwidth]{#2}
\vspace*{-6ex}
\end{minipage}}
\vspace*{-1ex}
 }
 
\begin{document}
 \tolerance=100000
 \topmargin -0.1in
\headsep 30pt
\footskip 40pt
\oddsidemargin 12pt
\evensidemargin -16pt
\textheight 8.5in
\textwidth 6.5in
\parindent 20pt
 
\def\baselinestretch{1.5}
\newcommand{\newc}{\newcommand}
\def\preprint{{preprint}}
\def\Ord{\lower .7ex\hbox{$\;\stackrel{\textstyle <}{\sim}\;$}}
\def\OOrd{\lower .7ex\hbox{$\;\stackrel{\textstyle >}{\sim}\;$}}
\def\cO#1{{\cal{O}}\left(#1\right)}
\newc{\order}{{\cal O}}
\def\lag             {{\cal L}}
\def\Lag             {{\cal L}}
\def\lum             {{\cal L}}
\def\R               {{\cal R}}
\def\Rsq             {{\cal R}^{\sq}}
\def\Rst             {{\cal R}^{\st}}
\def\Rsb             {{\cal R}^{\sb}}
\def\M               {{\cal M}}
\def\Oas             {{\cal O}(\alpha_{s})}
\def\Vcal            {{\cal V}}
\def\Wcal            {{\cal W}}
\newc{\be}{\begin{equation}}
\newc{\ee}{\end{equation}}
\newc{\br}{\begin{eqnarray}}
\newc{\er}{\end{eqnarray}}
\newc{\ba}{\begin{array}}
\newc{\ea}{\end{array}}
\newc{\bi}{\begin{itemize}}
\newc{\ei}{\end{itemize}}
\newc{\bn}{\begin{enumerate}}
\newc{\en}{\end{enumerate}}
\newc{\bc}{\begin{center}}
\newc{\ec}{\end{center}}
\newc{\ul}{\underline}
\newc{\ol}{\overline}
\newc{\ra}{\rightarrow}
\newc{\lra}{\longrightarrow}
\newc{\wt}{\widetilde}
\newc{\til}{\tilde}
\def\kr              {^{\dagger}}
\newc{\wh}{\widehat}
\newc{\ti}{\times}
\newc{\Dir}{\kern -6.4pt\Big{/}}
\newc{\Dirin}{\kern -10.4pt\Big{/}\kern 4.4pt}
\newc{\DDir}{\kern -10.6pt\Big{/}}
\newc{\DGir}{\kern -6.0pt\Big{/}}
\newc{\sig}{\sigma}
\newc{\sigmalstop}{\sig_{\lstoppair}}
\newc{\Sig}{\Sigma}  
\newc{\del}{\delta}
\newc{\Del}{\Delta}
\newc{\lam}{\lambda}
\newc{\Lam}{\Lambda}
\newc{\gam}{\gamma}
\newc{\Gam}{\Gamma}
\newc{\eps}{\epsilon}
\newc{\Eps}{\Epsilon}
\newc{\kap}{\kappa}
\newc{\Kap}{\Kappa}
\newc{\modulus}[1]{\left| #1 \right|}
\newc{\eq}[1]{(\ref{eq:#1})}
\newc{\eqs}[2]{(\ref{eq:#1},\ref{eq:#2})}
\newc{\etal}{{\it et al.}\ }
\newc{\ibid}{{\it ibid}.}
\newc{\ibidem}{{\it ibidem}.}
\newc{\eg}{{\it e.g.}\ }
\newc{\ie}{{\it i.e.}\ }
\def \viz{\emph{viz.}}
\def \etc{\emph{etc. }}
\newc{\nonum}{\nonumber}
\newc{\lab}[1]{\label{eq:#1}}
\newc{\dpr}[2]{({#1}\cdot{#2})}
\newc{\lt}{\stackrel{<}}
\newc{\gt}{\stackrel{>}}
\newc{\lsimeq}{\stackrel{<}{\sim}}
\newc{\gsimeq}{\stackrel{>}{\sim}}
\def\lsim{\buildrel{\scriptscriptstyle <}\over{\scriptscriptstyle\sim}}
\def\gsim{\buildrel{\scriptscriptstyle >}\over{\scriptscriptstyle\sim}}
\def\lapp{\mathrel{\rlap{\raise.5ex\hbox{$<$}}
                    {\lower.5ex\hbox{$\sim$}}}}
\def\gapp{\mathrel{\rlap{\raise.5ex\hbox{$>$}}
                    {\lower.5ex\hbox{$\sim$}}}}
\newc{\half}{\frac{1}{2}}
\newcommand {\nnc}        {{\overline{\mathrm N}_{95}}}
\newcommand {\dm}         {\Delta m}
\newcommand {\dM}         {\Delta M}
\def\bra{\langle}
\def\ket{\rangle}
\def\cO#1{{\cal{O}}\left(#1\right)}
\def \DM{{\Delta{m}}}
\newc{\bQ}{\ol{Q}}
\newc{\dota}{\dot{\alpha }}
\newc{\dotb}{\dot{\beta }}
\newc{\dotd}{\dot{\delta }}
\newc{\nindnt}{\noindent}

\newcommand{\medf}[2] {{\footnotesize{\frac{#1}{#2}} }}
\newcommand{\smaf}[2] {{\textstyle \frac{#1}{#2} }}
\def\onesq            {{\textstyle \frac{1}{\sqrt{2}} }}
\def\onehf            {{\textstyle \frac{1}{2} }}
\def\oneth            {{\textstyle \frac{1}{3} }}
\def\twoth            {{\textstyle \frac{2}{3} }}
\def\onefo            {{\textstyle \frac{1}{4} }}
\def\forth            {{\textstyle \frac{4}{3} }}

\newc{\matth}{\mathsurround=0pt}
\def\ML{\ifmmode{{\mathaccent"7E M}_L}
             \else{${\mathaccent"7E M}_L$}\fi}
\def\MR{\ifmmode{{\mathaccent"7E M}_R}
             \else{${\mathaccent"7E M}_R$}\fi}
\newcommand{\s}{\\ \vspace*{-3mm} }

\def \ud { {1 \over 2} }
\def \ut { {1 \over 3} }
\def \td { {3 \over 2} }
\newc{\mr}{\mathrm}
\def\dh {\partial }
\def \cs { cross-section }
\def \css { cross-sections }
\def \cm { centre of mass }
\def \cms { centre of mass energy }
\def \cc { coupling constant }
\def \ccs {coupling constants }
\def \gc {gauge coupling }
\def \gcc {gauge coupling constant }
\def \gccs {gauge coupling constants }
\def \yc {Yukawa coupling }
\def \ycc {Yukawa coupling constant }
\def \pp {{parameter }}
\def \pps {{parameters }} 
\def \ps {parameter space }
\def \pss {parameter spaces }
\def \vv {vice versa }

\newc{\siminf}{\mbox{$_{\sim}$ {\small {\hspace{-1.em}{$<$}}}    }}
\newc{\simsup}{\mbox{$_{\sim}$ {\small {\hspace{-1.em}{$>$}}}    }}


\newc {\Zboson}{{\mathrm Z}^{0}}
\newc{\thetaw}{\theta_W}
\newc{\mbot}{{m_b}}
\newc{\mtop}{{m_t}}
\newc{\sm}{${\cal {SM}}$}
\newc{\as}{\alpha_s}
\newc{\aem}{\alpha_{em}}
\def \PI{{\pi^{\pm}}}
\newc{\ppbar}{\mbox{$p\ol{p}$}}
\newc{\bbbar}{\mbox{$b\ol{b}$}}
\newc{\ccbar}{\mbox{$c\ol{c}$}}
\newc{\ttbar}{\mbox{$t\ol{t}$}}
\newc{\eebar}{\mbox{$e\ol{e}$}}
\newc{\zzero}{\mbox{$Z^0$}}
\def \gamz{\Gam_Z}
\newc{\wplus}{\mbox{$W^+$}}
\newc{\wminus}{\mbox{$W^-$}}
\newc{\ellp}{\ell^+}
\newc{\ellm}{\ell^-}
\newc{\elp}{\mbox{$e^+$}}
\newc{\elm}{\mbox{$e^-$}}
\newc{\elpm}{\mbox{$e^{\pm}$}}
\newc{\qbar}     {\mbox{$\ol{q}$}}
\def \ewgroup{SU(2)_L \otimes U(1)_Y}
\def \smgroup{SU(3)_C \otimes SU(2)_L \otimes U(1)_Y}
\def \smcolorem{SU(3)_C \otimes U(1)_{em}}

\def \SSM  {Supersymmetric Standard Model}
\def \poincare{Poincare$\acute{e}$}
\def \superspace{\emph{superspace}}
\def \sfs{\emph{superfields}}
\def \superpot{\emph{superpotential}}
\def \csf{\emph{chiral superfield}}
\def \csfs{\emph{chiral superfields}}
\def \vsf{\emph{vector superfield }}
\def \vsfs{\emph{vector superfields}}
\newc{\Ebar}{{\bar E}}
\newc{\Dbar}{{\bar D}}
\newc{\Ubar}{{\bar U}}
\newc{\susy}{{{SUSY}}}
\newc{\msusy}{{{M_{SUSY}}}}

\def\photino{\ifmmode{\mathaccent"7E \gam}\else{$\mathaccent"7E \gam$}\fi}
\def\taugluino{\ifmmode{\tau_{\mathaccent"7E g}}
             \else{$\tau_{\mathaccent"7E g}$}\fi}
\def\mphotino{\ifmmode{m_{\mathaccent"7E \gam}}
             \else{$m_{\mathaccent"7E \gam}$}\fi}
\newc{\gl}   {\mbox{$\wt{g}$}}
\newc{\mgl}  {\mbox{$m_{\gl}$}}
\def \charginopm{{\wt\chi}^{\pm}}
\def \mcharginopm{m_{\charginopm}}
\def \mchpmmin {\mcharginopm^{min}}
\def \chonep {{\wt\chi_1^+}}
\def \chone {{\wt\chi_1}}
\def \ch2p {{\wt\chi_2^+}}
\def \chonem {{\wt\chi_1^-}}
\def \ch2m {{\wt\chi_2^-}}
\def \chplus {{\wt\chi^+}}
\def \chminus {{\wt\chi^-}}
\def \chonip{{\wt\chi_i}^{+}}
\def \chonim{{\wt\chi_i}^{-}}
\def \chonipm{{\wt\chi_i}^{\pm}}
\def \chonjp{{\wt\chi_j}^{+}}
\def \chonjm{{\wt\chi_j}^{-}}
\def \chonjpm{{\wt\chi_j}^{\pm}}
\def \chonepm{{\wt\chi_1}^{\pm}}
\def \chonemp{{\wt\chi_1}^{\mp}}
\def \mchonepm{m_{\chonepm}}
\def \mchonemp{m_{\chonemp}}
\def \chtwopm{{\wt\chi_2}^{\pm}}
\def \mchtwopm{m_{\chtwopm}}
\newc{\dmchi}{\Delta m_{\wt\chi}}


\def \vlsp{\emph{VLSP}}
\def \lspi{\wt\chi_i^0}
\def \mlspi{m_{\lspi}}
\def \lspj{\wt\chi_j^0}
\def \mlspj{m_{\lspj}}
\def \lspone{\wt\chi_1^0}
\def \mlspone{m_{\lspone}}
\def \lsptwo{\wt\chi_2^0}
\def \mlsptwo{m_{\lsptwo}}
\def \lspthree{\wt\chi_3^0}
\def \mlspthree{m_{\lspthree}}
\def \lspfour{\wt\chi_4^0}
\def \mlspfour{m_{\lspfour}}


\newc{\sele}{\wt{\mathrm e}}
\newc{\sell}{\wt{\ell}}
\def \msell{m_{\sell}}
\def \slepone{\wt\ell_1}
\def \mslepone{m_{\slepone}}
\def \smuone{\wt\mu_1}
\def \msmuone{m_{\smuone}}
\def \stauone{\wt\tau_1}
\def \mstauone{m_{\stauone}}
\def \snu{\wt{\nu}}
\def \snutau{\wt{\nu}_{\tau}}
\def \msnu{m_{\snu}}
\def \msnumu{m_{\snu_{\mu}}}
\def \barsnu{\wt{\bar{\nu}}}
\def \barsnul{\barsnu_{\ell}}
\def \snul{\snu_{\ell}}
\def \mbarsnu{m_{\barsnu}}
\newc{\snue}     {\mbox{$ \wt{\nu_e}$}}
\newc{\smu}{\wt{\mu}}
\newc{\stau}{\wt{\tau}}
\newc {\nuL} {\wt{\nu}_L}
\newc {\nuR} {\wt{\nu}_R}
\newc {\snub} {\bar{\wt{\nu}}}
\newc {\eL} {\wt{e}_L}
\newc {\eR} {\wt{e}_R}
\def \slepl{\wt{l}_L}
\def \mslepl{m_{\slepl}}
\def \slepr{\wt{l}_R}
\def \mslepr{m_{\slepr}}
\def \stau{\wt\tau}
\def \mstau{m_{\stau}}
\def \slepton{\wt\ell}
\def \mslepton{m_{\slepton}}
\def \mlhiggs{m_{h^0}}

\def \xr{X_{r}}

\def \sfer{\wt{f}}
\def \msfer{m_{\sfer}}
\def \sq{\wt{q}}
\def \msq{m_{\sq}}
\def \msquleft{m_{\tilde{u_L}}}
\def \msqurht{m_{\tilde{u_R}}}
\def \sql{\wt{q}_L}
\def \msql{m_{\sql}}
\def \sqr{\wt{q}_R}
\def \msqr{m_{\sqr}}
\newc{\msqot}  {\mbox{$m_(\sq_{1,2} )$}}
\newc{\sqbar}    {\mbox{$\bar{\wt{q}}$}}
\newc{\ssb}      {\mbox{$\squark\ol{\squark}$}}
\newc {\qL} {\wt{q}_L}
\newc {\qR} {\wt{q}_R}
\newc {\uL} {\wt{u}_L}
\newc {\uR} {\wt{u}_R}
\def \ul{\wt{u}_L}
\def \mul{m_{\ul}}
\newc {\dL} {\wt{d}_L}
\newc {\dR} {\wt{d}_R}
\newc {\cL} {\wt{c}_L}
\newc {\cR} {\wt{c}_R}
\newc {\sL} {\wt{s}_L}
\newc {\sR} {\wt{s}_R}
\newc {\tL} {\wt{t}_L}
\newc {\tR} {\wt{t}_R}
\newc {\stb} {\ol{\wt{t}}_1}
\newc {\sbot} {\wt{b}_1}
\newc {\msbot} {m_{\sbot}}
\newc {\sbotb} {\ol{\wt{b}}_1}
\newc {\bL} {\wt{b}_L}
\newc {\bR} {\wt{b}_R}
\def \mul{m_{\wt{u}_L}}
\def \mur{m_{\wt{u}_R}}
\def \mdl{m_{\wt{d}_L}}
\def \mdr{m_{\wt{d}_R}}
\def \mcl{m_{\wt{c}_L}}
\def \charml{\wt{c}_L}
\def \mcr{m_{\wt{c}_R}}
\newc{\csquark}  {\mbox{$\wt{c}$}}
\newc{\csquarkl} {\mbox{$\wt{c}_L$}}
\newc{\mcsl}     {\mbox{$m(\csquarkl)$}}
\def \msl{m_{\wt{s}_L}}
\def \msr{m_{\wt{s}_R}}
\def \mbl{m_{\wt{b}_L}}
\def \mbr{m_{\wt{b}_R}}
\def \mtl{m_{\wt{t}_L}}
\def \mtr{m_{\wt{t}_R}}
\def \st{\wt{t}}
\def \mst{m_{\st}}
\newc {\stopl}         {\wt{\mathrm{t}}_{\mathrm L}}
\newc {\stopr}         {\wt{\mathrm{t}}_{\mathrm R}}
\newc {\stoppair}      {\wt{\mathrm{t}}_{1}
\bar{\wt{\mathrm{t}}}_{1}}
\def \lstop{\wt{t}_{1}}
\def \lstopbar{\lstop^*}
\def \hstop{\wt{t}_{2}}
\def \hstopbar{\hstop^*}
\def \mlstop{m_{\lstop}}
\def \mhstop{m_{\hstop}}
\def \lstoppair{\lstop\lstop^*}
\def \hstoppair{\hstop\hstop^*}
\newc{\tsquark}  {\mbox{$\wt{t}$}}
\newc{\ttb}      {\mbox{$\tsquark\ol{\tsquark}$}}
\newc{\ttbone}   {\mbox{$\tsquark_1\ol{\tsquark}_1$}}
\def \tsq {top squark }
\def \tsqs {top squarks }
\def \tsql {ligtest top squark }
\def \tsqh {heaviest top squark }
\newc{\mix}{\theta_{\wt t}}
\newc{\cost}{\cos{\theta_{\wt t}}}
\newc{\sint}{\sin{\theta_{\wt t}}}
\newc{\costloop}{\cos{\theta_{\wt t_{loop}}}}
\def \lsbot{\wt{b}_{1}}
\def \lsbotbar{\lsbot^*}
\def \hsbot{\wt{b}_{2}}
\def \hsbotbar{\hsbot^*}
\def \mlsbot{m_{\lsbot}}
\def \mhsbot{m_{\hsbot}}
\def \lsbotpair{\lsbot\lsbot^*}
\def \hsbotpair{\hsbot\hsbot^*}
\newc{\mixsbot}{\theta_{\wt b}}

\def \mhone{m_{h_1}}
\def \hup{{H_u}}
\def \hdn{{H_d}}
\newc{\tb}{\tan\beta}
\newc{\cb}{\cot\beta}
\newc{\vev}[1]{{\left\langle #1\right\rangle}}

\def \abot{A_{b}}
\def \atop{A_{t}}
\def \atau{A_{\tau}}
\newc{\mhalf}{m_{1/2}}
\newc{\mzero} {\mbox{$m_0$}}
\newc{\azero} {\mbox{$A_0$}}

\newc{\lb}{\lam}
\newc{\lbp}{\lam^{\prime}}
\newc{\lbpp}{\lam^{\prime\prime}}
\newc{\rpv}{{\not \!\! R_p}}
\newc{\rpvm}{{\not  R_p}}
\newc{\rp}{R_{p}}
\newc{\rpmssm}{{RPC MSSM}}
\newc{\rpvmssm}{{RPV MSSM}}


\newc{\sbyb}{S/$\sqrt B$}
\newc{\pelp}{\mbox{$e^+$}}
\newc{\pelm}{\mbox{$e^-$}}
\newc{\pelpm}{\mbox{$e^{\pm}$}}
\newc{\epem}{\mbox{$e^+e^-$}}
\newc{\lplm}{\mbox{$\ell^+\ell^-$}}
\def \branch{\emph{BR}}
\def \branche{\branch(\lstop\ra be^{+}\nu_e \lspone)\ti \branch(\lstop^{*}\ra \bar{b}q\bar{q^{\prime}}\lspone)}
\def \branchmu{\branch(\lstop\ra b\mu^{+}\nu_{\mu} \lspone)\ti \branch(\lstop^{*}\ra \bar{b}q\bar{q^{\prime}}\lspone)}
\def\Ecm{\ifmmode{E_{\mathrm{cm}}}\else{$E_{\mathrm{cm}}$}\fi}
\newc{\rts}{\sqrt{s}}
\newc{\rtshat}{\sqrt{\hat s}}
\newc{\gev}{\,GeV}
\newc{\mev}{~{\rm MeV}}
\newc{\tev}  {\mbox{$\;{\rm TeV}$}}
\newc{\gevc} {\mbox{$\;{\rm GeV}/c$}}
\newc{\gevcc}{\mbox{$\;{\rm GeV}/c^2$}}
\newc{\intlum}{\mbox{${ \int {\cal L} \; dt}$}}
\newc{\call}{{\cal L}}
\def \met  {\mbox{${E\!\!\!\!/_T}$}}
\def \cpv  {\mbox{${CP\!\!\!\!/}$}}
\newc{\ptmiss}{/ \hskip-7pt p_T}
\def \eslash{\not \! E}
\def \etslash{\not \! E_T }
\def \ptslash{\not \! p_T }
\newc{\PT}{\mbox{$p_T$}}
\newc{\ET}{\mbox{$E_T$}}
\newc{\dedx}{\mbox{${\rm d}E/{\rm d}x$}}
\newc{\ifb}{\mbox{${\rm fb}^{-1}$}}
\newc{\ipb}{\mbox{${\rm pb}^{-1}$}}
\newc{\pb}{~{\rm pb}}
\newc{\fb}{~{\rm fb}}
\newc{\ycut}{y_{\mathrm{cut}}}
\newc{\chis}{\mbox{$\chi^{2}$}}
\def \hadron{\emph{hadron}}
\def \nlc{\emph{NLC }}
\def \lhc{\emph{LHC }}
\def \cdf{\emph{CDF }}
\def\dzero{\emptyset}
\def \tevatron{\emph{Tevatron }}
\def \lep{\emph{LEP }}
\def \jets{\emph{jets }}
\def \jet(s){\emph{jet(s) }}

\def\Crs{stroke [] 0 setdash exch hpt sub exch vpt add hpt2 vpt2 neg V currentpoint stroke 
hpt2 neg 0 R hpt2 vpt2 V stroke}
\def\loopdk{\lstop \ra c \lspone}
\def\brloopdk{\branch(\loopdk)}
\def\fourdk{\lstop \ra b \lspone  f \bar f'}
\def\brfourdk{\branch(\fourdk)}
\def\fourdklep{\lstop \ra b \lspone  \ell \nu_{\ell}}
\def\fourdkhad{\lstop \ra b \lspone  q \bar q'}
\def\brfourdklep{\branch(\fourdklep)}
\def\brfourdkhad{\branch(\fourdkhad)}
\def\twodk{\lstop \ra b \chonep}
\def\brtwodk{\branch(\twodk)}
\def\threedkslep{\lstop \ra b \wt{\ell} \nu_{\ell}}
\def\brthreedkslep{\branch(\threedkslep)}
\def\threedksnu{\lstop \ra b \wt{\nu_{\ell}} \ell }
\def\brthreedksnu{\branch(\threedksnu) }
\def\threedklsp{\lstop \ra b W \lspone }
\def\brthreedklsp{\\branch(\threedklsp) }
\def\topdk{t \ra \lstop \lspone}
\def\rpvdk{\lstop \ra e^+ d}
\def\brrpvdk{\branch(\rpvdk)}
\def\fonec{f_{11c}} 
\newc{\mpl}{M_{\rm Pl}}
\newc{\mgut}{M_{GUT}}
\newc{\mw}{M_{W}}
\newc{\mweak}{M_{weak}}
\newc{\mz}{M_{Z}}

\newc{\OPALColl}   {OPAL Collaboration}
\newc{\ALEPHColl}  {ALEPH Collaboration}
\newc{\DELPHIColl} {DELPHI Collaboration}
\newc{\XLColl}     {L3 Collaboration}
\newc{\JADEColl}   {JADE Collaboration}
\newc{\CDFColl}    {CDF Collaboration}
\newc{\DXColl}     {D0 Collaboration}
\newc{\HXColl}     {H1 Collaboration}
\newc{\ZEUSColl}   {ZEUS Collaboration}
\newc{\LEPColl}    {LEP Collaboration}
\newc{\ATLASColl}  {ATLAS Collaboration}
\newc{\CMSColl}    {CMS Collaboration}
\newc{\UAColl}    {UA Collaboration}
\newc{\KAMLANDColl}{KamLAND Collaboration}
\newc{\IMBColl}    {IMB Collaboration}
\newc{\KAMIOColl}  {Kamiokande Collaboration}
\newc{\SKAMIOColl} {Super-Kamiokande Collaboration}
\newc{\SUDANTColl} {Soudan-2 Collaboration}
\newc{\MACROColl}  {MACRO Collaboration}
\newc{\GALLEXColl} {GALLEX Collaboration}
\newc{\GNOColl}    {GNO Collaboration}
\newc{\SAGEColl}  {SAGE Collaboration}
\newc{\SNOColl}  {SNO Collaboration}
\newc{\CHOOZColl}  {CHOOZ Collaboration}
\newc{\PDGColl}  {Particle Data Group Collaboration}

\def\issue(#1,#2,#3){{\bf #1}, #2 (#3)}
\def\ASTR(#1,#2,#3){Astropart.\ Phys. \issue(#1,#2,#3)}
\def\AJ(#1,#2,#3){Astrophysical.\ Jour. \issue(#1,#2,#3)}
\def\AJS(#1,#2,#3){Astrophys.\ J.\ Suppl. \issue(#1,#2,#3)}
\def\APP(#1,#2,#3){Acta.\ Phys.\ Pol. \issue(#1,#2,#3)}
\def\JCAP(#1,#2,#3){Journal\ XX. \issue(#1,#2,#3)} 
\def\SC(#1,#2,#3){Science \issue(#1,#2,#3)}
\def\PRD(#1,#2,#3){Phys.\ Rev.\ D \issue(#1,#2,#3)}
\def\PR(#1,#2,#3){Phys.\ Rev.\ \issue(#1,#2,#3)} 
\def\PRC(#1,#2,#3){Phys.\ Rev.\ C \issue(#1,#2,#3)}
\def\NPB(#1,#2,#3){Nucl.\ Phys.\ B \issue(#1,#2,#3)}
\def\NPPS(#1,#2,#3){Nucl.\ Phys.\ Proc. \ Suppl \issue(#1,#2,#3)}
\def\NJP(#1,#2,#3){New.\ J.\ Phys. \issue(#1,#2,#3)}
\def\JP(#1,#2,#3){J.\ Phys.\issue(#1,#2,#3)}
\def\PL(#1,#2,#3){Phys.\ Lett. \issue(#1,#2,#3)}
\def\PLB(#1,#2,#3){Phys.\ Lett.\ B  \issue(#1,#2,#3)}
\def\ZP(#1,#2,#3){Z.\ Phys. \issue(#1,#2,#3)}
\def\ZPC(#1,#2,#3){Z.\ Phys.\ C  \issue(#1,#2,#3)}
\def\PREP(#1,#2,#3){Phys.\ Rep. \issue(#1,#2,#3)}
\def\PRL(#1,#2,#3){Phys.\ Rev.\ Lett. \issue(#1,#2,#3)}
\def\MPL(#1,#2,#3){Mod.\ Phys.\ Lett. \issue(#1,#2,#3)}
\def\RMP(#1,#2,#3){Rev.\ Mod.\ Phys. \issue(#1,#2,#3)}
\def\SJNP(#1,#2,#3){Sov.\ J.\ Nucl.\ Phys. \issue(#1,#2,#3)}
\def\CPC(#1,#2,#3){Comp.\ Phys.\ Comm. \issue(#1,#2,#3)}
\def\IJMPA(#1,#2,#3){Int.\ J.\ Mod. \ Phys.\ A \issue(#1,#2,#3)}
\def\MPLA(#1,#2,#3){Mod.\ Phys.\ Lett.\ A \issue(#1,#2,#3)}
\def\PTP(#1,#2,#3){Prog.\ Theor.\ Phys. \issue(#1,#2,#3)}
\def\RMP(#1,#2,#3){Rev.\ Mod.\ Phys. \issue(#1,#2,#3)}
\def\NIMA(#1,#2,#3){Nucl.\ Instrum.\ Methods \ A \issue(#1,#2,#3)}
\def\JHEP(#1,#2,#3){J.\ High\ Energy\ Phys. \issue(#1,#2,#3)}
\def\EPJC(#1,#2,#3){Eur.\ Phys.\ J.\ C \issue(#1,#2,#3)}
\def\RPP (#1,#2,#3){Rept.\ Prog.\ Phys. \issue(#1,#2,#3)}
\def\PPNP(#1,#2,#3){ Prog.\ Part.\ Nucl.\ Phys. \issue(#1,#2,#3)}
\newc{\PRDR}[3]{{Phys. Rev. D} {\bf #1}, Rapid  Communications, #2 (#3)}

\vspace*{\fill}
\vspace{-1.5in}
\begin{flushright}
{\tt IISER/HEP/07/11}
\end{flushright}
\begin{center}
{\Large \bf Low mass
neutralino dark matter in mSUGRA and more general models in the light of LHC data}
  \vglue 0.4cm
  Nabanita Bhattacharyya\footnote{nabanita@iiserkol.ac.in},
  Arghya Choudhury\footnote{arghyac@iiserkol.ac.in} and
  Amitava Datta\footnote{adatta@iiserkol.ac.in}
      \vglue 0.1cm
          {\it 
	  Indian Institute of Science Education and Research, Kolkata, \\
          Mohanpur Campus, PO: BCKV Campus Main Office,\\
          Mohanpur - 741252, Nadia, West Bengal.\\
	  }
	  \end{center}
	  \vspace{.1cm}

\begin{abstract}
The $b \tau j$$\etslash$ signal at the ongoing LHC experiments is 
simulated with Pythia in the mSUGRA and other models of SUSY 
breaking. Special attention is given on the compatibility of this 
signature with the low mass neutralino dark matter (LMNDM) scenario
consistent with WMAP data. 
In the mSUGRA model the above signal as well as the LMNDM scenario are
strongly disfavored due to the constraints from the on going SUSY 
searches at the LHC. This tension, however, 
originates from the model dependent correlations among the parameters in 
the strong and electroweak sectors of mSUGRA. That there is no serious 
conflict between the LMNDM scenario and the LHC data is demonstrated 
by constructing generic phenomenological models such that the strong 
sector is
unconstrained or mildly constrained by the existing LHC data and  
parameters in the electroweak sector, unrelated to the strong sector,
yield DM relic density consistent with the WMAP data. 
The proposed models, fairly insensitive to the conventional SUSY 
searches in the jets + 
$\etslash$ and other channels, 
yield observable signal in the suggested channel for $\lum \gsim 1 \ifb$ 
of data. They are also consistent with the LMNDM scenario 
and can be tested by the direct dark matter search experiments in the 
near future. Some of these models can be 
realized by  non-universal scalar and gaugino masses at the GUT 
scale.
\end{abstract}

\newpage
\section{Introduction}

Proton - proton collisions at the LHC are now producing data at a center 
of mass energy ($\sqrt{s}$) = 7 TeV. There is no evidence of any new 
physics beyond the standard model (SM). However, it has already been 
shown by the ATLAS \cite{atlas} and CMS \cite{cms} collaborations that 
even with a small integrated luminosity ($\lum$) = 35 $pb^{-1}$, 
supersymmetry (SUSY) can be probed much beyond the existing limits on 
the sparticle masses obtained by the LEP \cite{lep} or Tevatron 
\cite{teva} experiments.

The negative results of new particle searches at the LHC have been 
interpreted in 
terms of the simplest gravity mediated SUSY breaking model - the minimal 
supergravity (mSUGRA) \cite{msugra} model- which has only five free 
parameters including soft SUSY breaking terms. These are $m_0$ (the 
common scalar mass), $m_{1/2}$ (the common gaugino mass), $A_0$ (the 
common trilinear coupling parameter), all given at the gauge coupling 
unification scale ($M_G \sim 2 \times 10^{16}$ GeV); the ratio of 
the Higgs vacuum expectation values at the electroweak scale 
namely tan$\beta$ and 
the sign of $\mu$. The magnitude of $\mu$ is determined by the radiative 
electroweak symmetry breaking (REWSB) condition \cite{rewsb}. The 
sparticle spectra and couplings at the electroweak scale are 
generated by renormalization group evolutions (RGE) of the above soft 
breaking masses and the coupling parameters. The non-observation of signal,
in particular in the jets + $\etslash$, channel leads to exclusion 
plots in the $m_0 - m_{1/2}$ 
plane. A large number of phenomenological analyses have also 
addressed the issue of SUSY search at LHC-7 TeV experiments 
\cite{tev7,nabanita1}.

In a hadron collider the dominant source of SUSY signals 
in the m $l$ + n jets + $\etslash$ channel is the pair
production of the strongly interacting sparticles - the 
squarks($\sq$) and gluinos($\gl$) - in various combinations. 
Throughout  this paper $l$ stands for $e$ and $\mu$ unless stated otherwise.
Thus the bounds from ATLAS \cite{atlas} and CMS \cite{cms} primarily exclude
some parameter space with relatively low  $m_0$ and $m_{1/2}$ 
which translates to certain combinations of squark $\sq$ and gluino $\gl$ 
masses \footnote{ The constraints become more severe due to the very 
recent $\lum$ = 1 $\ifb$ data as discussed briefly in 
Section 4.}. 
For example, the non-observation of the 0$l$ + jets + $\etslash$ signal 
implies that for nearly mass degenerate squarks and gluinos
$\msq \approx \mgl \geq 775$ GeV  (see the second paper
of \cite{atlas}). Here $\msq$ stands for the average mass of the  
L and R type squarks. 

On the other hand in several regions of the parameter space of the
minimal supersymmetric extension of the SM (MSSM) with conserved 
R-parity but without specific assumptions about the soft breaking 
parameters,
the  dark matter relic density \cite{dmrev,dmrev1} 
in the universe - low mass neutralino dark matter (LMNDM) 
in particular - essentially depends on the properties of the sparticles 
in 
the electroweak (EW) sector. This, e.g., is the case if the lightest 
neutralino, assumed to be lightest supersymmetric particle (LSP)
($\lspone$), is bino like and all squarks    
are beyond the reach of the ongoing LHC experiments.

However, due to the specific  correlations among the 
sparticle masses in mSUGRA, the above bounds on $\msq$ and $\mgl$ would 
also impose stringent indirect mass bounds on the EW sparticles. This
disfavors the LMNDM scenario.
The bound quoted above from ATLAS data, e.g., implies  
$\mslepr \gsim$ 398 and $\mlspone \gsim$ 125 {\footnote {Throughout this paper 
all masses, mass parameters and quantities having the dimension of mass are 
given in GeV unless stated otherwise.}}. These model 
dependent  bounds  imply that the masses of the 
sparticles belonging to the EW sector are far above the direct 
lower limits from LEP \cite{lep} and too large for realising the LMNDM 
scenario.          

The observed dark matter (DM)relic density ($\Omega h^2$) in the 
universe 
\cite{dmrev,dmrev1} has been precisely measured by the 
Wilkinson Microwave Anisotropy Probe (WMAP)
collaboration and is bounded by 0.09 $\leq \Omega h^2\leq$ 0.13 
\cite{wmap}.
A possible mechanism of production of relic density  in the above range
in the LMNDM scenario is annihilation of a bino like LSP pair or bulk 
annihilation \cite{dmrev1,bulk,plot}. 
It may be recalled that, if all strongly interacting sparticles are 
heavy, relatively low mass neutralinos and R type sleptons 
(super partners of $e_R$, $\mu_R$ and $\tau_R$) mainly contribute to this 
annihilation process. Coannihilation \cite{plot,coann} of a light 
neutralino and a nearly degenerate lighter stau mass eigenstate 
($\stau_1$) is another proposed mechanism for generating the observed 
relic density. The  allowed LMNDM scenarios in the 
mSUGRA model, with emphasis on  
the above two processes, have been delineated in the figures in 
\cite{debottam}
using parameter spaces  different from the conventional ones. In this 
paper we shall frequently refer to these figures. It seems that 
both the above processes are apparently in conflict with the recent LHC data. 
The
incompatibility of DM relic density production by slepton 
coannihilation and the data from  the 
LHC-7 TeV experiments have recently been noted in \cite{prannath}. 

The indirect 'exclusion' of a light  electroweak sector will have a
bearing on  direct detection of  DM \cite{direct} as well. The tension
between the constraints obtained by the ongoing LHC experiments and
the mSUGRA parameter space accessible to direct DM search experiments
by the XENON \cite{xenon} and the CDMS \cite{cdms} collaborations have
also been noted in the literature \cite{prannath,profumo}.
Several groups have also
reported on the prospect of constructing the mass of the neutralino
by such experiments. It is estimated that if $\mlspone \lsim$ 150,
then it might be possible to reconstruct this mass by measuring the
energy spectrum of the recoiling nuclear targets \cite{dmrev1,dmmass}.
The recoil energy spectrum is insensitive to higher neutralino masses.
Moreover the LMNDM scenario can be tested in an $e^+ - e^-$ collider
if the LSP mass is in the range 1 - 10 \cite{dreiner1}. However,
from the results of direct DM search and/or various constraints 
from collider and astrophysical experiments it has been claimed that 
the above mass range is disfavored \cite{feldman}.

In view of the above discussions it is worthwhile to critically 
reexamine the constraints from the ongoing LHC experiments and their 
impact on LMNDM scenarios. This will be taken up in a later section.  
The main conclusion is that in view of the uncertainties in the 
data, some parameter space with low $m_0$ and 
$m_{1/2}$ consistent with LMNDM cannot be conclusively ruled out. 
However, it 
must be admitted that there is a tension between the LHC data and the 
LMNDM scenario in the mSUGRA model.

We remind the reader that before  the advent of the bounds from the LHC, the 
bound $m_h > 114.4$ on the lighter Higgs scalar mass ($m_h$) 
from LEP \cite{hlim} tightly constrained the parameter space with 
low $m_0 -m_{1/2}$, a part of which coincides with the parameter space 
corresponding to 
the LMNDM scenario. These constraints are
particularly severe for low and intermediate values of tan$\beta$ 
\cite{CMSTDR}. 

It was recently emphasized in \cite{debottam,stark} that in order to 
revive
the parameter space consistent with the LMNDM scenario 
a moderate to large negative values of the trilinear coupling ($A_0$) 
is called 
for. This is particularly important if tan$\beta$ is not very large. In fact 
for $A_0$ = 0 and tan$\beta$ = 3 - a 
choice frequently employed by the LHC and Tevatron experiments - the
entire mSUGRA parameter space sensitive to the 35 $\ipb$ data is already excluded 
by the $m_h$ bound. For sizable negative values of $A_0$ the LMNDM 
scenario 
is realized for another reason. Here the lighter
stau mass eigenstate ($\stauone$) becomes significantly lighter than 
the selectron or the smuon even for moderate tan$\beta$ and LSP - 
$\stau_1$ coannihilation may occur efficiently. This happens 
for a minimum value of $m_{1/2}$ much lower
than the corresponding value for $A_0$ = 0 (see the figures depicting 
the parameter space allowed by WMAP data in \cite{debottam}).

This also gives 
rise to spectacular collider
signatures as the lighter chargino ($\chonepm$) - or the second lightest
neutralino ($\lsptwo$) dominantly decay into $\tau$ rich final states. 
Predictions for experiments 
at LHC-14 TeV \cite{debottam,nabanita} and LHC-7 TeV 
\cite{nabanita1} were worked out.  
In this paper we shall restrict 
ourselves to non-zero values of $A_0$ only.

It is, however, well-known that the 0$l$ + jets + $\etslash$ signal
is fairly insensitive to the choice of $A_0$ and tan$\beta$ (see the 
second paper in \cite{atlas}). Throughout
this paper we shall assume that the bounds obtained by the ATLAS and the 
CMS collaborations in this channel for fixed choices of these two 
parameters are valid for other choices as well.

Another  important consequence of non-zero $A_0$ is that the lighter top
squark mass eigenstate $\lstop$ may be significantly lighter than
the other squarks and could be copiously produced at the ongoing LHC
experiments. They may come from two dominant sources; i) Direct
$\lstop - \lstop^*$ pair production and ii) $\gl \ra \lstop \bar{t}$, if
kinematically allowed.

Our next task is to identify a signal which unlike the jets + $\etslash$
final state is sensitive to $|A_0|$. If $\lstop$ is not the next 
lightest super particle, then the decay mode $\lstop \ra b \chonep$ may
be its main decay channel resulting in final states rich in b-jets.
If tan$\beta$ is small (say, 5) then the electroweak gauginos will 
dominantly 
decay leptonically into $e$, $\mu$ or $\tau$ channels with approximately
equal probability. 
A viable signal in this case would be $ b l j$$\etslash$ \cite{arghya}. 

For moderate or large tan$\beta$ and non-zero $A_0$ on the other hand 
the above EW gauginos decay dominantly into final states involving 
$\tau$'s leading to very characteristic collider signals 
\cite{nabanita1,debottam,nabanita}. The price to be paid for $\tau$ 
tagging efficiency may be adequately compensated by the large BRs of the 
EW gaugino decays into $\tau$ rich final states. The main purpose of 
this paper is to study the observability $ b \tau j$$\etslash$ at the 
ongoing LHC experiments with emphasis on the LMNDM scenario. Here
$j$ is the number of jets in the signal and different choices of this 
variable will be considered.  
Occasionally, however, we shall also revisit the $b l j$$\etslash$ 
signal \cite{arghya}.

As discussed  above there is indeed a tension between the realisation  
LMNDM in mSUGRA and the preliminary data from the 
LHC. Should the experimental constraints be strengthened in future, the 
tension will further intensify. In view of this we propose a few generic 
models which 
are either mildly constrained or unconstrained by the current LHC data 
and are consistent with the low mass neutralino DM scenario. The main 
point is 
that the LHC data is sensitive to the masses of the strongly interacting 
sparticles while the realization of LMNDM in these models  DM hinges on 
the properties of the electroweak sector. Thus if the two sectors are 
uncorrelated the above tension will cease to exist. These models are 
generic in the sense that their viability depends on certain mass 
hierarchies among the strongly interacting sparticles rather than on some
specific choices of the masses. The important correlations among 
different mass hierarchies in  SUSY models and the corresponding collider 
signatures have been emphasized in the literature\cite{hierarchies}. 
The parameters in the electroweak sector
can be chosen independently. In fact all models where the sparticle 
masses in the EW sector consistent with the corresponding LEP limits,
derived without assuming mSUGRA as the underlying model, are allowed  
in principle. These models are phenomenological in nature 
\footnote{In spirit these models are similar to the simplified 
phenomenological model considered by the ATLAS collaboration with only squarks of the 
first two generation, the gluinos and the LSP within the reach of the 7 
TeV  run (see Fig. 2 in the second paper  of \cite{atlas}). }
although we shall
comment on  theoretical motivations wherever possible. Finally we 
shall discuss the possibility of testing these models by the ongoing  
LHC experiments.

The plan of the paper is as follows. In Section 2 we shall 
mainly concentrate on the phenomenological models 
unconstrained by the LHC data corresponding to $\lum$ = 35 $\ipb$
above and assess  the prospect of observing the $ b \tau j 
\etslash$ signatures. Special emphasis will 
be given on the LMNDM scenarios. 
In Section 3 we shall examine the $b \tau j$$\etslash$ signal and 
occasionally the $ b l j$$\etslash$ signal in the
mSUGRA model using the above data and comment on the viability of realizing  the LMNDM 
scenarios in view of the uncertainties in the LHC constraints. 
In Section 4 the analyses of Section 2 and 3 are updated in the light of the
recent $\lum = 1 \ifb$ data.
Our 
conclusions will be summarized in Section 5.

\section{SUSY signatures and LMNDM in generic 
models}

The simplest generic model compatible with all LHC data 
accumulated so far ($\lum$ = 35 $\ipb$) would be one
with all strongly interacting sparticles beyond the reach of 7 TeV 
experiments while all sparticles in the electroweak sector are light.
Unfortunately the earlier simulations in the context of the 14 TeV run
indicate that any signature in the current experiments at lower energy 
is not likely. Thus for a model with non-trivial signatures at this 
stage of the
LHC experiment one needs at least one relatively light strongly 
interacting sparticle.

In the first model with modest values of tan $\beta$, only the third 
generation of squarks 
and the sparticles in the EW sector are assumed to be within the reach 
of the early phases of 7 TeV run. The $b l j$$\etslash$ signal has already 
been studied in \cite{arghya} in such a phenomenological scenario .  
The large mixing in the top squark mass matrix producing a light
mass eigenstate  provides  a qualitative justification. For small or 
moderate tan$\beta$ the b-squark mass eigenstates will be much heavier. 
Of course a large trilinear soft breaking term $A_t$ is
needed for the above  mixing. It  will also yield $m_h$ compatible with 
the LEP bound through radiative corrections. 

The current LHC data
hardly constrain this model since events from direct low mass $\lstop - 
\lstop^*$ pair production  have too little $\etslash$ or $m_{eff}$ 
\cite{arghya} to survive
the strong cuts currently implemented by the ATLAS and CMS experiments
for SUSY search.
Dedicated searches with softer cuts are called for. 

In this model the first two generations 
of squarks and the gluinos are assumed to be beyond the reach of the
early stages of the on going LHC experiments due to some yet unknown 
soft breaking mechanism. We further assume for the sake of simplicity that 
$m_0$ is the common mass of the squarks belonging
to the third generation, the sleptons and the two neutral Higgs bosons. 
Similarly $m_{1/2}$ controls the masses of the electroweak gauginos 
only. If this partially constrained spectrum yields observable signal 
over a reasonably 
large parameter space it is obvious that more will be available in a 
totally unconstrained MSSM. In fact if the unification of the electroweak
gaugino masses at $M_G$ is relaxed, the mass of the LSP DM candidate
can be even lower \cite{dreiner}.

We shall fix tan$\beta$ = 10 but take $A_0$ as a variable. 
Again for simplicity the magnitude of $A_0$ is
restricted  to be less than 1 TeV. While computing the spectrum
we have checked that  no charge-colour breaking minimum 
of the scalar potential \cite{ccb} occurs. For each point the minimum
allowed $A_0$ is determined by the $m_h$ bound from LEP \cite{hlim}.   

The radiative corrections to
the lightest Higgs boson mass ($m_h$) involve some theoretical 
uncertainties (see, for example, \cite{debottam, arghya} for a brief 
discussion 
and references to the original works). In view of these uncertainties, 
if the computed Higgs mass is $m_h > 110$ \footnote{See footnote 4.} 
for a point in the parameter space, that point will be regarded as 
acceptable although the computed mass is somewhat smaller than the 
direct bound from the direct searches at LEP 
\cite{hlim}. Throughout this work the pole mass of the top (bottom) 
quark will be taken as $m_t (m_b)=$ 173 (4.25) \cite{topmass}.
We shall assume that the masses of the 
lighter chargino, all the sleptons except the tau mass eigenstates
and the third generation squarks are heavier than 100. This is 
basically a simplified form of the LEP limits. The lighter $\stau$
mass eigenstate is assumed to be heavier than 82 as required
by the LEP data.

The light sparticle masses and decay branching ratios (BRs)are generated 
by SUSPECT \cite{suspect} and Sdecay \cite{sdecay} and $\mu$ is fixed by 
the REWSB condition \cite{rewsb}.

In the parameter space of interest for the on going sparticle searches  
at the LHC the two body decays $\lstop \ra b 
\chonep$ occurs with almost 100\% branching ratio (BR). Moreover
in bulk of the parameter space the decays  
$\chonep \ra \stau_1 \nu,\quad \snu \tau$ dominate, leading to the 
$b \tau j$$\etslash$ signature. When the latter decays become subdominant
or is kinematically suppressed
the decay into a real W ($\chonep \ra \lspone W$) or a virtual W 
($\chonep \ra \lspone W^* \ra \lspone f \bar{f^{\prime}})$ open 
up yielding the $b l j$$\etslash$ signature \cite{arghya}.    
  
We have generated $\lstop \lstop^*$ pair events at $E_{CM} =
7$ TeV using Pythia \cite{pythia}. The  signal  $b \tau j$$\etslash$ 
has been simulated using the following 
procedure. Initial and final state radiation,
decay, hadronization, fragmentation and jet formation are implemented
following the standard procedures in Pythia.

In this paper all leading order (LO) signal cross-sections have been 
computed by CalcHEP \cite{calchep} unless otherwise stated. For any two 
body final state (except for QCD processes) with identical particles or 
sparticles both the renormalization and the factorization scales are 
taken as, $\mu_R = \mu_F = M$, where $M$ is the mass of the particle or 
sparticle concerned. For two unequal masses the scales are taken to 
be the average of the two. For QCD events the scales have been chosen to be 
equal to $\sqrt {\hat s}$ which is the energy in the parton CM frame, 
and the cross-section is computed by Pythia. All LO cross-sections are 
computed using CTEQ5L parton density functions (PDFs) \cite{cteql}.

We have considered the backgrounds from  $t \bar t$, QCD events and
$W$ + $n$-$jets$ events, where $W$  decays into all channels.
$t \bar t$ events are generated using Pythia and the LO 
cross-section has been taken from CalcHEP which is 85.5 pb.
QCD processes are generated by Pythia in different $\hat p_T$ bin :
$25 \le \hat p_T \le 400$, $400 \le \hat p_T \le 1000$ and 
$1000 \le \hat p_T \le 2000$ , where $\hat p_T$ is defined in
the rest frame of the  parton collision.
The main contribution comes from the low $\hat p_T$ bin,
which has a cross-section of $\sim 7.7 E+07$ pb.
However, for other bins ($ 400 < \hat p_T < 1000$ and
$1000 < \hat p_T < 2000$), the background events are
negligible.

For $W$ + $n$-$jets$ events we have generated events with $n=0,1$ and 
$2$ at the parton level using ALPGEN (v 2.13) \cite{alpgen}.
We have generated these events subjected to the condition
that $P_T^j > 20$, $\Delta R(j,j) \ge 0.3$ and
$\vert \eta \vert \le 4.5$. These partonic events have been fed to 
Pythia for parton showering, hadronization, fragmentation and decays etc.

The next to leading order (NLO) cross-sections for stop-stop pair production 
have been computed by PROSPINO \cite{prospino} using the CTEQ5M PDFs.
The K-factors are computed by comparing with the LO cross-section. The LO 
cross-sections from PROSPINO agree well with  CalcHEP for the same choice of 
the scales. 

The NLO background cross-sections are not known for some 
backgrounds - in particular for the QCD processes.  
For computing the significance of the signal we conservatively multiply 
the total  LO background by an overall factor of 2.

We have used the toy calorimeter simulation (PYCELL) provided in Pythia
with the following settings.

\begin{itemize}
\item The calorimeter coverage is $\vert \eta \vert < 4.5$. 
The segmentation is given by $\Delta \eta \times \Delta \phi = 0.09 
\times 0.09$ which
resembles a generic LHC detector.

\item A cone algorithm with $\Delta$ R$ = \sqrt {\Delta\eta^2 + 
\Delta\phi^2}= 0.5 $ has been used for jet finding.

\item E$^{\mathrm{jet}}_{\mathrm{T,min}} = 30$ and jets are ordered
in E$\mathrm{_T}$.
\end{itemize}

The signal has been selected as follows:

\underline{Lepton Veto:}

Leptons $(l=e,\mu)$ are selected with $P \mathrm{_T \ge 10}$
and $\vert\eta \vert < 2.4$. For lepton-jet isolation
we require $\Delta R(l,j) > 0.5$. For the sake of simplicity
the detection efficiency of $e$ and $\mu$ are assumed to be $ 100 \%$.
Events with isolated leptons are rejected.

\underline{$b$- jet identification:}

We have tagged $b$-jets in our analysis by the following procedure.
A jet with $|\eta|< 2.5$ corresponding to the coverage of tracking 
detectors
matching with a $B$-hadron of decay length $> 0.9$ mm
has been marked $tagged$.
This criteria ensures that single $b$-jet tagging efficiency
(i.e., the ratio of tagged $b$-jets and the number of taggable $b$-jets)
$\epsilon_b \approx 0.5$ in $t \bar t$ events.

\underline{$\tau$- jet identification:}

Taus are identified through their hadronic decays producing
narrow jets with 1 or 3 tracks pointing to the jets.
We have defined a narrow signal cone
of size $\Delta R_S= 0.1$ and an isolation cone of size $\Delta R_I= 0.4$
around the calorimetric jet axis. We then require 1 or 3 charged tracks
inside the signal cone with $|\eta_{track}|< 2.5$ and
$P_T > 3$ for the hardest track.
We further require that there are no other charged tracks with $P_T > 1$
inside the isolation cone to ensure tracker isolation. 
 
 The following cuts will be call {\it Set 1}. These cuts ensure stop rich signal events
 while rejecting the background efficiently :

\begin{itemize}
\item We have selected events with one {\it tagged} $b$ jet ({\it cut 1.1}).
\item We have selected events with one {\it tagged} $\tau $ jet ({\it cut 1.2}).
\item We have rejected events with isolated lepton ({\it cut 1.3}).
\item Events with missing transverse energy ($\etslash) \ge 70$ are selected ({\it cut 1.4}).
\item We have also demanded events with $P_T$ {\it tagged} $\tau$ jet $\ge 40$ ({\it cut 1.5}).
\bc
or
\ec
We have demanded events with $P_T$ {\it tagged} $b$ jet $\le 50$ 
({\it cut 1$^{\prime}$.5}). This cut is particularly useful if the 
mass difference between the $\lstop$ and the $\chonepm$ is small. 
It also rejects the $t \bar{t}$ background efficiently.
\end{itemize}
The set of cuts which includes {\it cut 1$^{\prime}$.5} will be referred to
as {\it Set 1$^{\prime}$}.

Our results are summarized in Fig. 1.
In the green (online) or small crosshatched (red (online) or big crosshatched)) 
region the $b \tau j$$\etslash$ signal 
can be observed for an integrated luminosity of $\lum$ = 1 (5) $\ifb$.
For observability we simply require $S/\sqrt{B}>$ 5, where S (B) is the 
number of signal (background) events. We find that $\mlstop$
upto 235 (280) can be probed by this signal with $\lum$ = 1 (5) $\ifb$.

In the blue (online) or hatched region the decays  $\chonep \ra \stau_1 \nu_{\tau}, \snutau \tau$
are either phase space
suppressed or are kinematically forbidden. But the $b l j$$\etslash$ 
signal becomes observable with the cuts of {\it Set 1} proposed in 
\cite{arghya} for a suitable $A_0$. In the blue (online) or hatched region $\mlstop$ varies
between 130 - 190 for suitable choices of $A_0$. Beyond a certain $m_0$
(approximately 480), $A_0 >$ 1 TeV will be required to make $\mlstop$
sufficiently light for an observable signal.

To get a feeling for the relative size of the signal and the SM backgrounds 
we consider four benchmark points BP1 (110,170,-540,10), BP2 (110,190,-640,10), 
BP3 (130,200,-550,10), BP4 (150,230,-745,10)
which  yield observable signals at  $\lum =$ 1 $\ifb$ (see Fig. 1). The 
quantities in the bracket are $m_0, m_{1/2}$,
$A_0$ and tan$\beta$. The stop mass at the four points are 143.6, 
165.0, 215.6 and 208.2 respectively. At all four benchmark points 
BR($\lstop \ra b \chonep$) and BR($\chonep 
\ra \stau_1^+ \nu$) are maximal to a very good approximation.

The 0$l$ + jets + $\etslash$ or any other signal which gets 
contribution from all squark-gluino production is not particularly 
sensitive to the presence or absence of a light stop. However 
the $b \tau j$$\etslash$ signal proposed here stems from 
$\lstop - \lstop^*$ pair production alone. Hence it sensitively 
depends on $\mlstop$ and ,consequently, on $A_0$.
The quoted $A_0$ gives the largest signal in each case. However, for each point a 
range of $A_0$ values leads to observable signals. In this range  
the $\lstop$-$\chonep$ mass difference corresponding to $|A_0|_{max}$  
is too small for producing  a taggable 
b-jet and the signal becomes weak. In contrast at $|A_0|_{min}$, the 
$\lstop - \lstop^*$ production is 
too suppressed - due to a relatively heavy $\lstop$- to produce a viable 
signal. This is true for all points in Fig. 1. For example with
$m_0$ = 90, $m_{1/2}$ = 180, the signal is observed in  the range $|A_0|=$ 
370 - 550. For e.g., $A_0$ = -370, -500 and -550 observable signals occur 
for $\lum = $ 5 $\ifb $, $\lum =$ 1 $\ifb $ and $\lum =$ 5 $\ifb $ respectively.

The LO cross-sections after different cuts for the four 
benchmark points and different SM backgrounds
are presented in Table 1.

\begin{table}[htb]
\begin{center}\
\begin{tabular}{|c|c|c|c|c||c|c|c|c|}
\hline
             & \multicolumn{4}{c|}{Signal} & \multicolumn{4}{c|}{Background} \\
\hline
              			& BP1   & BP2  & BP3 & BP 4 & $t \bar t$ &     QCD  & $W + 1j$    & $W + 2j$  \\
$\sigma$ (pb) 			&38.9   &19.2  &4.56 &5.54&    85.5    & $7.7 \times10^7$  &  $1.43 \times10^4$   &  5200     \\
\hline
\hline

{\it cut 1.1} 			&4.6251     &3.5190 &2.2193  & 1.6781 &42.1780  &$6.454\times10^5$  &44.0797  &48.1231    \\
\hline
{\it cut 1.2} 			&1.0696     &0.6797 &0.7327  &0.5540&3.8987  &$6.426\times10^4$     &2.7408   &2.9982   \\
\hline
{\it cut 1.3} 			&0.8581     &0.5443 &0.5365  &0.4164&2.8266  &$6.348\times10^4$     &2.4310   &2.6816    \\
\hline
{\it cut 1.4} 			&0.2899     &0.1887 &0.2469  &0.1689&0.7301  &22.1066       &0.1787   &0.3067    \\
\hline
{\it cut 1.5} 			&0.2075     &0.1262 &0.1802  &0.1278& 0.5318 &0.0102        &0.1191   &0.2126    \\
or       			&              &       &        &      &        &              &         &         \\
{\it cut 1$^{\prime}$.5}	& 0.2067    &0.1273 &0.0784  &0.1045& 0.1855 &0.0001        &0.0595   &0.0940    \\
\hline
$S/\sqrt B$ 			&	    & 	    & 	     &	    &   	&	&	  &	    \\
After {\it cut 1.5}	 	& 8.4	    &5.1    &7.3     &5.2   &	 	&	&	  &	    \\
After {\it cut 1$^{\prime}$.5}	&13.4	    &8.3    &5.1     &6.8   &	 	&	&	  &	    \\
\hline

\hline
\end{tabular}
\end{center}
\caption{The LO  cross-sections (including efficiency) for the signal 
corresponding to  BP1 - BP4 in the phenomenological model ( section 
2, paragraphs 2 - 4) and the SM 
backgrounds after 
{\it Set 1 } or {\it Set 1$^{\prime}$} of cuts. 
The last row gives the significance of the signal for $\lum =$ 1 $\ifb$ 
on the basis of NLO cross-sections (see text). }
\end{table}

The last row gives the significance of the signal for $\lum =$ 1 
$\ifb$ for the two sets of cuts. The significances, however, are 
computed on the basis of the NLO cross-sections as discussed above. For 
the range of $\mlstop$ relevant here the K-factor varies rather slowly.
We have multiplied the LO signal cross-sections by the average value 
which is approximately 1.7. 
As stated above the total SM background is multiplied by an overall 
factor of 2.

It follows that larger significance is obtained in some regions by
using {\it cut 1.5} while in others {\it cut 1$^{\prime}$.5} is more 
appropriate.
The {\it cut 1.5} will be more effective in the region where
the $P^{\tau}_T$ is harder. This is the case for BP3 
as a result of the relatively large mass difference between $\stau_1$  
and $\lspone$. 
On the other hand for regions where the difference between $m_{\lstop}$
and $m_{\chonep}$ is less than 50, {\it cut 1$^{\prime}$.5} will be more 
effective which is the case for  BP1, BP2 and BP4. 
\begin{figure}[!htb]
\begin{center}
\includegraphics[angle =270, width=1.0\textwidth]{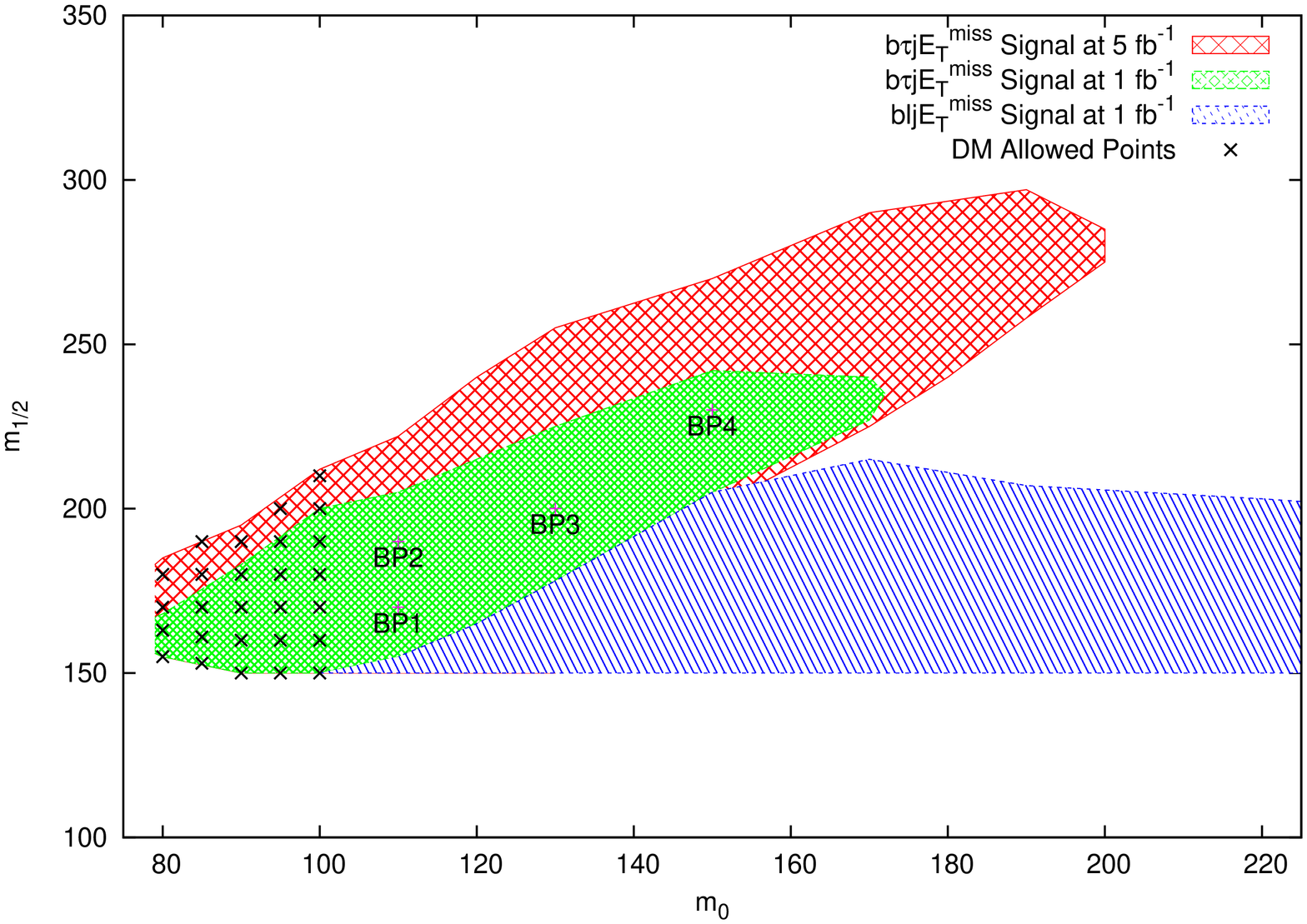}
\end{center}
\caption{Regions of $m_0$ - $m_{1/2}$ plane which can be probed by direct 
production of 
$\lstop \lstop^*$ pairs for 1 $\leq$ $\lum$ $\leq$ 5 $\ifb$ (see 
Section 2, paragraph 2 -4 for the details of the underlying  
phenomenological model). The SM backgrounds are in Table 1.}
\end{figure}


In Fig. 1 each point marked with a cross is consistent 
with the DM relic density data \footnote{We remind the reader that
the parameters $m_0$ and $m_{\half}$ used here are different from 
the common scalar and gaugino masses in mSUGRA. We have defined these
parameters in the present context in Section 2, paragraph 4.}.      
In Figure 1 we also 
identify the parameter space sensitive to the proposed signal.
Therefore, in this reach plot we do not restrict ourselves to points 
allowed by the WMAP data only. In fact at many points in the delineated 
parameter 
space the computed $\Omega h^2$ violates even the WMAP upper bound on the
relic density. If a 
signal is indeed observed at
any of these points, it would indicate that the lightest neutralino 
though
stable in the time scale of a collider event may be cosmologically 
unstable.
A tiny R-parity violation induced by higher dimensional operators, 
for example, may induce neutralino decays at the cosmological scale. In 
such cases the observed relic density must come from some other 
non-neutralino sources.  

In \cite{debottam} it was shown that there are two generic regions 
corresponding to LMNDM. There is a region where neutralino pair 
annihilation via R-type light slepton exchange or bulk annihilation 
produces the observed relic density. The tension between the computed 
lighter Higgs scalar mass ($m_h$)and the corresponding experimental 
lower bound is softened by non-zero but moderate negative values of 
$A_0$ (200 - 300) and the uncertainties in the computed $m_h$. Here the 
$b \tau j$$\etslash$ signal is observable since the $\stau_1$ - 
$\lspone$ mass difference is relatively large. Hence taggable 
$\tau$-jets arise from the decay $\stau_1 \ra \tau \lspone$. 

In the 
other regions consistent with LMNDM the lighter $\stau$ mass eigenstate 
is much lighter due to larger values of $|A_0|$ ($\order$ (1 TEV)). As a 
result $\stau_1$ - $\lspone$ coannihilation along with bulk annihialion 
produce the LMNDM. However, due to 
the small $\stau_1$ - LSP mass difference the $b \tau j$$\etslash$ 
signal from stop pair production may not be viable.

The next scenario we consider is one with two strongly interacting 
sparticles within the reach of the first phase of the LHC experiment 
along with electroweak sparticles. It is assumed that the production of 
the lighter top squark and the gluinos is the primary source of the LHC 
signatures. The other strongly interacting sparticles are assumed to be 
beyond the reach of the LHC-7 TeV run. The masses of all strongly 
interacting sparticles are taken to be independent of the electroweak 
sector. For simplicity the masses of the sparticles and other relevant 
parameters in the EW sector are chosen as in mSUGRA. One is thus free to 
choose any LMNDM scenario. This will be referred to as the Light Stop 
Gluino (LSG) scenario. 

We shall first consider a phenomenological model 
with $\mlstop$ and $\mgl$ as unrelated parameters at the weak scale.
We shall also consider a mSUGRA type scenario with non-universal
boundary conditions at the GUT scale which leads to the LSG scenario.

However, the parameter space in the LSG scenario is already  constrained 
to some extent by the 
35 $\ipb$ LHC data. The ATLAS collaboration has analysed
the jets + $\etslash$ signal using four sets of cuts referred to as
A, B, C, D \cite{atlas}. The corresponding lower limits on the 
production cross-sections of all strongly interacting sparticles 
including the efficiencies are 
1.3, 0.35, 1.1 and 0.11 pb respectively. These limits are converted into
constraints in the $m_0 - m_{1/2}$ plane by computing the NLO 
cross-section by PROSPINO and the efficiencies by Pythia.  

In mSUGRA the set D consisting of very hard cuts is very potent for 
obtaining new mass bounds. In the LSG scenario, however, this set kills 
the 
signal in any LSG model consistent with the Tevatron bounds on $\mgl$ 
for heavy squarks \cite{teva}. When each point in the parameter space is 
required to pass all the above cut sets, set A and set C turns out to be 
most effective in obtaining limits in the LSG scenario. The resulting 
bounds are $\mgl \geq$ 320-330 for $\mlstop$ = 150 - 300. The 
insensitivity of the bound on $\mgl$ to $\mlstop$ clearly indicates that 
the hard 
cuts employed by the ATLAS group in isolating the jets + $\etslash$ 
signal eliminates all events from stop pair production. The bound essentially
comes from gluino pair production followed by the decay of each gluino into
top-stop pairs. Hence the canonical  signals with strong cuts
are not suitable for revealing the presence of a light stop. Interesting
contribution to the signal may, however, originate from gluino decays.

For computing the above  limits we have set the masses of the squarks (also 
sleptons) of L and R type belonging to the first two generations at 1.5 
TeV. We have treated $\mtl, \mtr, \mbr, A_t, A_b$ as well as $\mgl$ as 
free parameters.  We have chosen the following parameters for the 
electroweak sector at the weak scale as $M_1$ =60, $M_2$ =125, 
$m_{\stau_L}$ =155, $m_{\stau_R}$ =116, $A_{\tau} =-615$, tan$\beta$ = 
10, $\mu$ = 348 . This choice of parameters gives $\mcharginopm = 122,$ 
$\mlspone= 59$, $\mstauone = 109$, $m_{\wt{\nu}_{\tau_1}} = 142$ and 
compatible with the WMAP data. The above limits mildly depend on the 
variation of the parameters in the EW sector unless one makes very 
specific choices such that the decay pattern of the EW gauginos are 
drastically different ( e.g., they all decay leptonically with 100\% 
BR).

SUSY searches at the Tevatron can also potentially constrain the LSG 
scenario. We have not done the analysis in this paper. As a cautious 
approach we have only considered $\mgl \gsim$ 400 which corresponds to 
the bound on $\mgl$ if all squarks are heavy and is stronger than the 
limits discussed above.

We next compute the $b \tau j$$\etslash$ signal for different $\mlstop$
and $\mgl$ in the LSG scenario. The other parameters are fixed as  
follows: $m_{\tilde q_{L/R}(1,2)}$ = $\mslepl$ = $\mslepr$ = 1.5 TeV, 
$m_{\chonepm}$ = 152.6, $m_{\lspone}$ = 79.1, $m_{\wt \tau_1}$ = 144.1. We 
take wino like $\chonepm$ and bino like $\lspone$. The electroweak 
sector is chosen such that the resulting DM relic density is consistent
with WMAP data.

\begin{table}[!htb]
\begin{center}
\begin{tabular}{|c|c|c|c||c|c|c|}
\hline
        Points             & \multicolumn{3}{c|}{$\lstop \lstop^*$} & \multicolumn{3}{c|}{ LSG scenario} \\
	\hline
$m_{\lstop} $-$m_{\wt g}$  &{\it Cut Set 1}&{\it Cut Set 2}&{\it Cut Set 3}& {\it Cut Set 1}&{\it Cut Set 2}	&{\it Cut Set 3}		\\
       \hline
209 - 504		   &359.0(8.6)	&270.3(8.8)	&3.8(0.9)	&463.4(11.1)	&280.5(9.1)	&28.7(7.2)	\\	
	\hline
250 - 443   		   &204(4.9**)	&136.7(4.4**)	&6.8(1.7)	&285.3(6.8)	&144.7(4.7**)	&25.4(6.3)	\\	  
	\hline
250 - 517 		   &209(5.0)	&140.7(4.5**)	&6.6(1.6)	&239.2(5.7)	&139.2(4.5**)	&20.1(5.0)	\\
	\hline
       \end{tabular}
       \end{center}

\caption {Number of $b \tau j \etslash$ events from pure $\lstop 
\lstop^*$ production (columns 2 - 4) and that in 
the phenomenological LSG scenario (columns 5 - 7) for three sets
of cuts (see text). 
Numbers in the brackets are the significance of the signal 
for 
$\lum =$ 1 $\ifb$ using NLO cross-section and the entries 
marked with **  indicate that the signal is observable for 1 $< \lum 
\leq$ 5 $\ifb$. }

\end{table}

We analyse the signal with 3 sets of cuts and the results are 
summarized in table 2. The {\it Cut Set 1} and the corresponding SM 
background have already been given. Both stop and gluino production can 
contribute to the signal. If the stop is relatively heavy and the gluino
is relatively light the size of the overall BSM signal may improve
considerably to the case where only $\lstop$ of the same mass is 
present.

The {\it Cut Set 2} has 
two cuts in addition to {\it Cut Set 1}:
\begin{itemize}
\item $ N_{Central-jet} \le 4 $ ( where the central jets have 
$\vert \eta \vert \le$ 2.5) .
\item $ M_{eff} \le 500 $ ( where $M_{eff}= |\met| + \Sigma_{i}|P_T^{j_i}| + \Sigma_{i}|P_T^{l_i}|$ ($l_i = e,\mu$)).
\end{itemize}
The corresponding SM background is 0.4744 pb. 
After this set of cuts the gluino
induced events in the sample are drastically reduced. As discussed in 
\cite{arghya}
this sample can be used for studying the properties of the light stop
squarks. Moreover, if the signal comes from stop pair production alone
the fraction of events surviving this cut is much larger than that if
the signal stems from both $\lstop$ and gluino induced events. This 
distinction 
, however, becomes unclear as $\mgl$ increases. 

The {\it Cut Set 3} is designed to remove the pure stop induced events. 
It consists of the following stronger cuts :

\begin{itemize}

\item We have selected events with one {\it tagged} $b$ jet  ({\it cut 3.1}).

\item We have selected events with one {\it tagged} $\tau $ jet ({\it cut 3.2}).
\item We have rejected events with isolated leptons ({\it cut 3.3}).
\item Events with at least 6 central jets are selected, where central 
jets are defined as pycell jets with $\vert \eta \vert \le$ 2.5 ({\it cut 3.4}).

\item Events with missing transverse energy ($\etslash) \ge 160 $ 
are selected ({\it cut 3.5}).

\end{itemize}

\begin{table}[!htb]
\begin{center}\
\begin{tabular}{|c|c|c|c|c|}
\hline
              & $t \bar t$ &     QCD  & $W + 1j$    & $W + 2j$  \\
\hline
{\it cut 3.1}	 &42.1780  	 &$6.454\times10^5$	&44.0797  &48.1231   	 \\
\hline
{\it cut 3.2}	 &3.8987	 &$6.426\times10^4$	&2.7408	  &2.9982   	\\
\hline
{\it cut 3.3} 	 &2.8266	 &$6.348\times10^4$ 	&2.4310	  &2.6816   	 \\
\hline
{\it cut 3.4} 	 &0.4454	 &0.0256 	&0.0476	  &0.2275   	 \\
\hline
{\it cut 3.5} 	 &0.0076	 &0.0004 	&-	  &-   		 \\
\hline

\end{tabular}
\end{center}
\caption{The LO cross-section (including efficiency) of SM backgrounds 
after 
the  {\it Set 3} of cuts.}
\end{table}
 
Only The gluino induced events survive and depending on the gluino mass 
may 
give an observable signal. The significance of the signal for
$\lum =$ 1 $\ifb $ for each case is given in parentheses in  
Table 2. The corresponding SM backgrounds are presented in Table 3. 
For the gluino mass range in Table 2 the K factor varies
between 1.6 and 1.7.

The LSG scenario can be realized in gravity mediated SUSY breaking
with non-universal masses at $M_G$. All scalar superpartners 
squarks belonging to 
the first two generations  are assumed to have masses beyond the reach
of the ongoing runs at the LHC. The squarks belonging to the third 
generation are assumed to have much smaller masses. At $M_G$ the $\stau$ 
mass is assumed to be even smaller so that the LMNDM can be eventually
realized. Qualitatively the non-universal masses for the third 
generation can be generated by the running of the corresponding 
soft breaking parameters, from 
the SUSY breaking scale down to $M_G$ \cite{nonunisca}, 
by effective mass terms induced at $M_G$ by some yet unknown flavour 
dependent interactions above the GUT scale etc, etc. The 
phenomenology of these models has attained due attention 
\cite{nonuniscapheno}.   

The gluino mass ($M_3$) is assumed to be smaller than the other 
gaugino masses at $M_G$. For simplicity it is assumed that $M_1=M_2$ 
at this scale. They are chosen such that they are compatible with a
LMNDM scenario. The motivation for non-universal gaugino masses have 
already been discussed by several authors \cite{nonunigauge}. It arises 
if a GUT nonsinglet chiral superfield couples to the gauge kinetic 
function and the hierarchy among the non-universal gaugino masses 
depends on the representations of the GUT group to which the chiral 
superfield belongs. In practice linear combinations of such chiral 
superfields may couple, 
making the prediction of the above mass hierarchy rather difficult.   
The phenomenology non-universal gaugino masses has also been 
discussed extensively \cite{nonunigaugepheno}. Models with both the above 
non-universalities have also been constructed \cite{gennonuni} 
and the resulting phenomenology were analyzed \cite{nonunipheno}. 

Finally in order to get the magnitude of $\mu$ consistent with a LMNDM
scenario with bino like LSP, 
we have to introduce non-universal soft breaking Higgs masses 
$m_{H_u}$ and $m_{H_d}$ at $M_G$ \cite{nonunihiggspheno}.   
          
For illustration we have chosen the following spectrum at $M_G$ which 
is consistent with a LMNDM scenario satisfying the WMAP 
constraints, where the relic density is 
produced by stau - LSP coannihilation:
$m_0(1,2)$ = 1.5 TeV, 
$m_{H_u}$= 300, $m_{H_d}$= 500, $M_1 = M_2 $= 240, $A_t = A_b = A_{\tau} $ = $-500$, tan$\beta$ = 10.

The parameter $m_0(3)(\wt t)$, the common mass for the third 
generation of squarks ($\wt Q_L$, $\wt t_R$, 
$\wt b_R$) and  the gluino mass $M_3$ at $M_G$ are the variables.  
The common mass  $m_0(3)(\stau)$ of the third generation of sleptons ($\wt \tau_L $, $\wt \tau_R$, $\wt \nu_{\tau}$) 
is mildly varied in 
a small  range to obtain a LMNDM scenario. 
The variation of the soft breaking terms at $M_G$ and the resulting 
physical masses of the relevant sparticles are presented in Table 4.

\begin{table}[!htb]
\begin{center}
\begin{tabular}{|c|c|c||c|c|c|c|c||c||c|}
\hline

$M_3$	& $m_0(3)$  &$m_0(3)$  &$\mlstop$&$\mchonepm$&$\mstauone$ &$\mlspone $& $\mgl $ & S1	    & S3	\\
	& $(\st)$& $(\stau)$&	&	&	 &	& 	 		&({\it Cut Set 1})   &({\it Cut Set 3})	\\
\hline
	&460	&112 	&214	&187	&106.3	&98.0	&429		&329.8(7.9)	&35.2(8.8)	\\
150	&520	&117 	&288	&190	&107.4	&98.6	&435		&196.3(4.7**)	&20.1(5.0)	\\
	&580	&121	&351	&192	&107.7	&98.9 	&438		&136.1(3.3**)	&18.7(4.7**)	\\

\hline
170	&460	&114	&250	&188	&106.6	&98.0 	&475		&235.6(5.6)	&23.6(5.9)	\\
\hline
200	&460	&117	&302	&189	&107.4	&98.0 	&547		&75.8(1.8)	&10.6(2.6**)	\\
\hline

       \end{tabular}
       \end{center}
          \caption{Some representative  non-universal input parameters 
at $M_G$ leading to the LSG scenario 
and the corresponding sparticle spectra. The EW sector yields DM relic 
density consistent with WMAP data. Last two 
columns give number of $b \tau j \etslash$  events 
(using NLO cross-sections) under cut 
{\it Set 1} and {\it Set 3} for $\lum =$ 1 $\ifb$. Significance 
for each case is given in the bracket and entries marked with ** indicate 
that signal is observable for 1 $<$ $\lum$ $\leq$  5 $\ifb$. }
          \end{table}

In the phenomenological LSG scenario $\mlstop$ is a weak scale input and 
$A_t$ is not important. In the LSG scenario obtained from  non-universal 
GUT scale boundary conditions, a smaller $\mlstop$ contributes more to 
the size of the $b \tau j$$\etslash$ signal obtained with softer cuts (see 
the last but one columns of table 4 and section 4). Hence $A_t$ at $M_G$ is
an important parameter (see the inputs for table 4).

The prospect of observing the $b \tau j$$\etslash$ signal at the ongoing 
LHC experiments in the above LSG scenario is also summarized in Table 4. 
It follows that for a fixed $M_3 = 150 (\mgl \approx 
435)$, $\mlstop \leq$  350 can be probed with $\lum$ a little more than
1 $\ifb$.         

The $b \tau j \etslash$ signal, which arises when the chargino
decays into $\tau$ rich final states with large BR, has 
not so far been searched  by 
the LHC collaborations. The ATLAS collaboration has already searched for 
the $b l j \etslash$ signal arising from the decay of stop pairs using 
the 35 $\ipb$ data (see the third paper in ref. 1). Such pairs 
may either be produced directly or via two body decay of gluino pairs 
produced at the LHC. Apparently the experiment takes into account both 
possibilities and excludes $\mgl$ below 530 - 540 for $\mlstop$ = 125 - 
300. However, the insensitivity of the gluino mass limit on $\mlstop$ 
indicates that the events from directly produced stop pairs are 
eliminated by the selection procedure and effectively their limits arise 
from the gluino induced processes. As already shown in \cite{arghya} and 
also in this paper directly produced light top squarks can also 
contribute to the signal under a different search strategy based on 
softer cuts. The possibility that only lighter stop squark pairs have 
been produced at the LHC yielding either the $b l j \etslash$ or 
$b \tau j \etslash$ signature still remains open.  

We first employ relatively soft cuts so that
the search is sensitive to both direct stop pair
production and
gluino pair production followed by the decay of each gluino into top-stop
pairs. This enhances the size of the overall SUSY signal especially for
relatively heavy stop ($\mlstop \geq 250$) and light gluino. 
As demonstrated above one can 
design cuts which can separate the signal stemming from the two
different production channels.

As already noted the simplest model practically unconstrained by the LHC data
would of course be the one in which all strongly 
interacting sparticles are beyond the reach of the LHC experiments at 7 TeV. 
This model can be motivated by introducing non-universal gaugino masses in 
a mSUGRA like framework  by requiring $M_3 >> m_{1/2}, m_0$ at $M_G$ 
, where $M_3$ is the gluino mass and $m_{1/2}$ is the common mass of 
the $SU(2)$ and $U(1)$ gauginos \cite{adnabanita}. The rest of the 
sparticle masses may be determined by mSUGRA like parameters $m_0,
M_1 = M_2 = m_{1/2}, A_0$, tan$\beta$ and sign ($\mu$). The parameter
$m_0$, assumed to be much smaller than $M_3$ at $M_G$, 
controls the scalar masses in the electroweak sector but has very
little impact on the masses of the strongly interacting sparticles. 
The latter
parameters can be chosen such that they are consistent with the LMNDM
scenario.   

Unfortunately the signal will not be easy to observe at the on going
LHC experiments. The slepton pair production \cite{slepton} and the clean 
trilepton signal \cite{trilepton} from chargino ($\chonepm$)-second 
neutralino ($\lsptwo$) pair production would be the most distinctive
signatures of this model. It is, 
however, well known that 
even for the 14 TeV experiment the reach in the trilepton channel is 
rather modest \cite{cmstrilepton}. This does not suggest an exciting 
prospect for the 7 TeV run. At 14 TeV the reach can be improved by 
including the 1$l$ + 2 $\tau$ or 2$l + 1\tau$ events arising from 
the decays of $\chonepm$ and $\lsptwo$ \cite{adnabanita}. The reach
for slepton search is also rather limited at 14 TeV \cite{cmsslepton} 

These 
signals, however, 
have not been studied for the on going run nor is the prospect of 
slepton search at 7 TeV  known. In this paper we did not pursue 
this model any further.

\section{SUSY signatures and LMNDM  in mSUGRA 
in view of LHC data with $\lum$ = 35 $\ipb$ }

In this section we consider the prospect of observing the $b \tau j$$\etslash$ 
signal at LHC-7 TeV and realizing the LMNDM scenario in mSUGRA.
However, the ATLAS collaboration has obtained strong constraints on 
the mSUGRA parameter space \cite{atlas} which should be taken into 
account. We remind
the reader that the ATLAS results were obtained for $A_0$ = 0 and 
tan$\beta$ = 3 - a choice forbidden by the direct bound on $m_h$ from LEP. 
However, in the same paper they noted that the 
limits are rather insensitive to the variation of the above two 
parameters. We shall, therefore, assume that these limits are valid in more general
models with different $A_0$ and tan$\beta$ values.

\begin{figure}[!htb]
\begin{center}
\includegraphics[angle =270, width=1.0\textwidth]{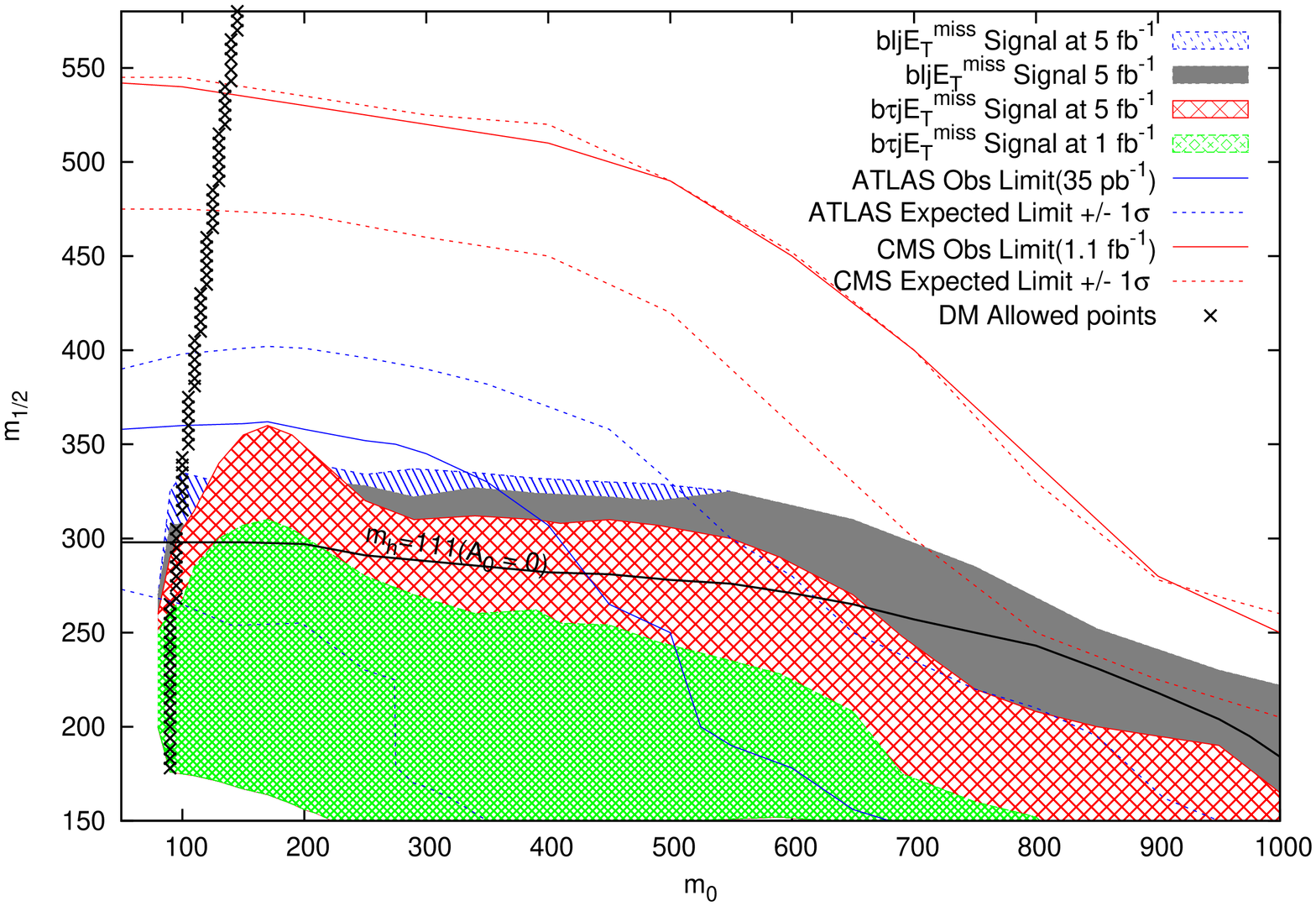}
\end{center}
\caption{The green (online) or small crosshatched (red (online) or big crosshatched)) 
region of $m_0$ - $m_{1/2}$ plane in mSUGRA can be probed by the $b \tau 
j$$\etslash$ signal from 
all squark-gluino events using {\it Set 3} of cuts with $\lum =$ 1 (5) $\ifb$. 
The $b l j$$\etslash$ signal probes the grey (blue (online) or hatched) region with the cuts proposed in 
\cite{arghya} (cut {\it Set 4} introduced in this paper) for 1$< \lum \leq$ 5 $\ifb$. Here
$A_0$ = -600 and tan$\beta$ = 10, sign($\mu$) $>$ 0 (see text for the details).}
\end{figure}

Our results are summarized in Fig. 2. This figure corresponds to 
$A_0$ = $-600$ and tan$\beta = $ 10.
The observed ATLAS limits (reproduced in Fig. 2) are 
rather strong and strongly disfavors the LMNDM scenario. However, these
limits were obtained with the strongest set of cuts used by them (signal region  D
in \cite{atlas}) resulting in  2 observed events against a SM 
expectation of $2.5 \pm 1.0 ^{1.0}_{-0.4}\pm 0.2$. An upward fluctuation of the 
observed number, which is certainly a distinct possibility, would relax the limits significantly. 
Thus the variation of the expected limit, which
is very close to the observed limit in this case, in response to $\pm 
1 \sigma$ fluctuation of the SM expectation is perhaps a more realistic 
description of our present knowledge. The band resulting from the above
fluctuations as given in \cite{atlas} is also included in Fig. 2.

We also present the points (denoted by the x mark) in the low 
$m_0$-$m_{1/2}$ region allowed 
by the DM relic density data. The relic density is mainly produced by
stau-LSP coannihilation with some contribution from bulk annihilation.
The observed limits disfavor most of the  region consistent with the 
LMNDM scenario. In fact $\mlspone \leq  143 $ is disfavored.  
However, if the uncertainties in the limits are taken into
account $\lspone$ with smaller masses can not be the excluded with certainty 
as DM candidates. It is also to
be noted that the computed $m_h$ for $A_0$ = 0 also puts pressure on 
the low $m_0$ - $m_{1/2}$ region even after  
theoretical uncertainties in the computation is taken into account.  

The {\it Cut Set 3} introduced in Section 2 is used for 
estimating the signal and the background (see Table 3). 
In the green (online) or small crosshatched region the signal is observable with $\lum =$ 1 $\ifb$. 
No signal from the points consistent with the observed relic density
is expected for DM allowed points even after considering the 
uncertainties in the ATLAS data. For 1 $\ifb < \lum <$ 5 $\ifb$ the 
signal is observable over a much larger region of the parameter space 
(the red (online) or big crosshatched region). 
A few points allowed by the relic density data yield observable signals
(see fig. 2).

We next comment briefly on the $b l j$$\etslash$ signal. In addition to
the {\it Set 3} of cuts defined in \cite{arghya} we introduce a new set of cuts 
({\it Set 4}) given below.

\begin{itemize}
\item We have selected events with one {\it tagged} $b$ jet.
\item We have selected events with one isolated lepton .
\item Events with at least 6 central jets are selected, where central
jets are defined as pycell jets with $\vert \eta \vert \le$ 2.5.
\item Events with missing transverse energy ($\etslash) \ge 160 $ are
selected.
\end{itemize}
The former set 
gives a better reach in $m_{1/2}$ for $m_0 <$ 550 while the second set
is more effective for larger $m_0$ for $\lum =$ 5 $\ifb$. This is 
illustrated in Fig. 2 by the grey and the blue (online) or hatched regions respectively.

Recently several groups have reexamined the sensitivity of the on going 
XENON100 experiment in the light of the constraints on the mSUGRA model 
imposed by the LHC 35 $\ipb$ data. The Figure 2(L) of \cite{profumo}, for 
example, indicates that the parameter space sensitive to the above experiment 
is
not compatible with the LHC constraints. However, if the sensitivity of
XENON experiment is improved by a factor of 10, which may be achieved by
the end of 2012, a much larger region of the mSUGRA parameter space may be
accessible to direct search experiments. Discovery of neutralino DM 
by direct search in the
near future would, therefore, suggest models beyond mSUGRA like the ones 
discussed in this paper.  
\section{ The generic models, the mSUGRA model and the LMNDM scenario revisited in view of
the $\lum$ = 1 $fb^{-1}$ data }

After submitting this paper to the arXiv and the journal.  
the constraints on the mSUGRA model from LHC data with $\lum =$ 1 $\ifb$ 
were announced. In this section we discuss briefly the impact 
of the new data on our main results. The following discussion is based 
on the results presented in the Lepton-Photon conference, 2011 
\cite{leptonphoton} and in the CMS paper \cite{cmsnew}.

Of course the results for the $b\tau j$$\etslash$ signal 
from stop pair production alone in the first part of Section 2 remain unaltered. 
As already discussed, none of the CMS and ATLAS searches employing 
very hard $\etslash$ cuts are sensitive to either 
the $blj$$\etslash$ signal discussed in \cite{arghya} 
\footnote {The above comment is also applicable to the 
negative search in $blj$$\etslash$ channel by the ATLAS collaboration 
\cite{atlasbnew} briefly reviewed below.} 
or the $b\tau j$$\etslash$ channel discussed in this paper. 
A dedicated search for relatively low mass top squarks using softer cuts as 
out lined in Section 2 is called for. Also the search for final states 
with one tagged b-jet and one $\tau$ jet has not been reported so far 
by the LHC collaborations.

The stronger constraints from the new data in the jets + $\etslash$ 
channel will increase the lower bounds on $\mgl$ in the LSG scenario 
presented in Section 2. However, the details of the cross section 
limits similar to those presented by the ATLAS group for the 35 $\ifb$ data 
used in obtaining our bounds, are not yet available for the $\lum$ = 1 $\ifb$ data. 
We, therefore, use the following procedure.

From the ATLAS analysis in the jets + $\etslash$ channel based on 165 
$pb^{-1}$ of 2011 data \cite{atlas0lnew} the limit on 
gluino mass is 500 for heavy squarks($m_{\tilde q}$ = 1250). The CMS 
analysis in the same channel using 1.14 $\ifb$ of 2011 data \cite{cmsnew} 
puts a  stronger limit on the gluino mass. For heavy squarks with average 
mass 1.5 TeV (2 TeV), the lower limit on gluino mass is 570 (550). 
In the Lepton-Photon conference, 2011 \cite{leptonphoton}  
ATLAS results in the same channel was presented 
for 1.04 $\ifb$ of 2011 data. For heavy squarks gluino mass 
below 600 are excluded. 
   
Strictly speaking the above limits are not directly applicable in the 
LSG scenario. In this case a gluino decays via the cascade $\wt g \ra 
\lstop t$ $\ra(b\charginopm)(bW^{\pm})$, yielding the signal whereas CMS 
and ATLAS consider gluino pair production in the limit where all 
squarks are heavy, so that each gluino decays via 3 body 
modes($q\qbar\lspone$, $q\qbar\lsptwo$ or $q q^\prime \charginopm$). 
Jets are, therefore, likely to be softer in our case on the average. 
Hence the hard cuts on jet $p_T$ employed by the LHC collaborations 
are expected to give somewhat weaker limit on $\mgl$ in the LSG 
scenario. As a reasonable guess we have considered $\mgl$ $\ge$ 550 in the 
LSG scenario. Results for $\mgl \leq 500$ are already available in 
Section 2.

\begin{table}[!htb]
\begin{center}
\begin{tabular}{|c|c||c|c|c||c|c|c|}
\hline
 \multicolumn{2}{|c||}{Points}& \multicolumn{3}{c||}{$\lstop \lstop^*$} & \multicolumn{3}{c|}{ LSG scenario} \\
\hline
$m_{\lstop} $&$m_{\wt g}$  &{\it Cut Set 1}&{\it Cut Set 2}&{\it Cut Set 3} & {\it Cut Set 1}	&{\it Cut Set 2}&{\it Cut Set 3}\\
       \hline
    & 550	           &		&		&		&394.4(9.4)	&285.8(9.3)	&17.4(4.3**)	\\
209 & 700		   &362.9(8.7)	&272.3(8.9)	&4.5(1.1)	&378.3(9.0)	&275.4(9.0)	&10.3(2.5**)	\\
    & 850		   &		&		&		&374.2(8.9)	&270.3(8.8)	&8.8(2.2)**	\\
	\hline
	\hline
    & 550   		   &		&		&		&230.7(5.4)	&138.2(4.4**)	&16.3(4.0**)	\\	  
250 & 650 		   &207.6(5.0)	&136.0(4.4)	&6.8(1.6)	&210.8(5.1)	&136.5(4.4**)	&10.2(2.5**)	\\
    & 750 		   &		&		&		&204.3(4.9**)	&132.1(4.2**)	&8.3(2.1)	\\
	\hline
       \end{tabular}
       \end{center}

\caption {Notations, conventions and input parameters are the same as 
Table 
2.}

\end
{table}
The ATLAS group has also updated their search for the  
$blj$$\etslash$ signal \cite{atlasbnew} in the LSG 
scenario. Using 1.03 $\ifb$ data they have excluded
gluino masses below 500-520  $\mlstop$ in the 
range 125 - 300. This limit, however, is also not applicable to 
the analysis in this paper based on 
a different chargino decay mode (BR($\charginopm$ $\ra \stauone 
\nu_{\tau}$)=100 $\%$). 
\begin{table}[!htb]
\begin{center}
\begin{tabular}{|c|c|c||c|c|c|c|c||c||c|}
\hline

$M_3$	& $m_0(3)$  &$m_0(3)$  &$\mlstop$&$\mchonepm$&$\mstauone$ &$\mlspone $& $\mgl $ & S1	    & S3	\\
	& $(\st)$& $(\stau)$&	&	&	 &	& 	 		&({\it Cut Set 1})   &({\it Cut Set 3})	\\
\hline
\hline
210	&365	&112	&229	&186	&106	&97 	&567		&228.7(5.5)	&16.7(4.1**)	\\
	&420	&114	&283	&188	&105	&98 	&569		&101.9(2.4**)	&12.2(3.0**)	\\
\hline
225	&330	&112	&231	&185	&106	&97	&600		&218.8(5.2)	&13.9(3.5**)	\\	
\hline
250	&250	&111	&235	&184	&106	&97	&657		&201.3(4.8**)	&9.8(2.4**)	\\	
\hline

       \end{tabular}
       \end{center}
          \caption{Notations, conventions and input parameters are 
the same as in Table 4. }
          \end{table}

We present in Table 5 the observability of the signal in the 
phenomenological LSG model where $\mlstop$ and $\mgl$ are unrelated 
parameter. This table illustrates the search prospect for $\mgl \geq $ 
550. The notations, input parameters consistent with a LMNDM scenario 
and cuts are the same as in Table 2. 
The corresponding results for the scenario with nonuniversal boundary 
conditions at the GUT scale (see Section 2) are presented in Table 6.  
It may also be noted that if $\lum$ = 10 $\ifb$ is accumulated in the 
LHC 7 TeV run then mass reach in this channel will be considerably 
improved.  For example if $\mgl$ = 550 then $\mlstop \sim 315 $ can be probed 
with {\it Cut Set 1}.

The new data from LHC based on $\lum$ = 1 $\ifb$, however, strongly 
disfavor the LMNDM scenario in the mSUGRA model irrespective of the 
experimental uncertainties. We superimpose on Figure 2 the region 
excluded by the CMS collaboration \cite{cmsnew}. It is readily seen that 
for low $m_0$, neutralino - stau coannihilation is the only viable 
mechanism for DM relic density production provided 
$\mlspone \geq 215 $.

Even if the issue of neutralino dark matter is set aside, the $b\tau j$$\etslash$ 
is unobservable in the mSUGRA model even for $\lum$ = 5 $\ifb$ 
(see the recent CMS constraint reproduced in figure 2). 
The $blj$$\etslash$ signal is observable in a small corner of the parameter space 
thanks to the experimental uncertainties as reflected by the CMS 
expected limit $\pm 1 \sigma$ band.
   

\section{Summary and conclusions}

The data from the on going experiments at the LHC has put a question
mark on the viability of a
substantial region of the mSUGRA parameter space corresponding to
low mass sparticles. This leads to a tension between the data 
and
the low mass neutralino dark matter (LMNDM) scenario where  
neutralino annihilation and/ or neutralino - slepton coannihilation can
produce DM relic densities consistent with the WMAP data. Moreover,
large regions of the parameter space sensitive to the ongoing 
experiments for direct DM search  or searches in the near future are 
also 
under pressure.

However, the LHC experiments are primarily sensitive to the masses
of the strongly interacting sparticles - the squarks and the gluinos.
In contrast in  typical LMNDM scenarios the relic density may  
depend entirely 
on the properties of the sparticles in the electroweak sector. Thus the
above tension is an artifact of the model dependent correlations among
the soft breaking masses in the strong and the electroweak sectors of
mSUGRA. Any model in which the  masses in the two sectors are 
independent parameters could be free from this tension.                  

In view of this we propose a few generic models which are unconstrained
or mildly constrained by the LHC data. These models are generic in the 
sense that their acceptability depends on certain mass hierarchies in 
the
strong sector and not on specific mass values. The electroweak sector 
is
assumed to be independent of the strong sector. In fact we only assume 
that the LSP is bino like, tan $\beta$ has intermediate values  and all 
parameters in the EW  sector 
are consistent with the corresponding LEP 
limits  provided such limits are not based on mSUGRA dependent 
assumptions.         

In the first phenomenological model under consideration all strongly 
interacting 
sparticles except for the lighter top squark are assumed to be beyond 
the reach of the experiments at LHC-7 TeV (see Section 2). This squark 
and the electroweak sparticles are the only sources of SUSY signals. For 
simplicity we assume that the masses of the relatively light sparticles 
are correlated 
as in mSUGRA. Our conclusions  will obviously hold in a 
more general framework. In this case the trilinear soft breaking parameter 
($A_0$) must be non-zero. This ensures consistency with the bound on 
the Higgs mass from LEP \cite{debottam}. Moreover, for a given $m_0$, 
LSP-$\stauone$ coannihilation may produce the observed relic density 
for 
values of $m_{1/2}$ significantly lower than  that for $A_0 = 0$ 
\cite{debottam}. This also facilitates the LMNDM scenario.  

The conventional SUSY signals with hard cuts on $\etslash$ or $m_{eff}$
are not viable in this model. However, a fairly large region of the
parameter space can be probed by the $b \tau j$$\etslash$ signal (see
Table 1 and the reach plot Fig. 1) using the search strategy sketched in 
this paper. It is estimated that $\mlstop \leq$ 
235 (280) 
can be probed with $\lum$ = 1 (5) $\ifb$.    
A part of this parameter space is consistent with the WMAP
data where the DM relic density is produced by bulk annihilation
as shown in Fig 1. However, this result is subject to certain 
simplifying assumptions (see Section 2 paragraph 4). In an unconstrained 
MSSM the 
signal will be compatible with a much larger parameter space consistent 
with the LMNDM scenario.  

The complementary signal $b l j \etslash$ signal 
proposed in \cite{arghya} may be useful if the charginos do not 
dominantly decay into tau rich final states. The possibility that 
signals from stop pair production only are already buried in the LHC 
data is still open.    

We next focus on a model (the LSG scenario)in which only $\lstop$ and 
$\gl$ but no other
strongly interacting sparticle, are within the striking range of the 
ongoing LHC experiments. As before the EW sector is assumed to be 
independent of the strong sector and is taken to be consistent with the 
WMAP data. This model is only mildly constrained 
by the LHC 
data for $\lum = 35~\ipb$(see Table 2 for some representative 
examples). 
The $b \tau j$$\etslash$ signal can probe a fairly large region of the 
parameter 
space (Table 2). Moreover, using additional 
cuts the stop induced events can be separated from the gluino induced 
ones (see Table 2). The sample thus separated can be used for 
reconstruction of the properties of $\lstop$. 

The very recent data for $\lum = 1 \ifb$, however, make the 
lower limit on $\mgl$ more stringent ($\mgl \geq 550$) for heavy squarks
(see Section 4).
Thus the contribution of gluino pair production to the signal reduce
significantly. Still the stop induced and gluino induced contributions 
can be separated by 
suitable cuts as illustrated in (Table 5).  

The LSG scenario can be realized in  a mSUGRA type model with 
nonuniversal 
boundary conditions at $M_G$ (see Table 4 for sample results for $\lum$ 
= 35 $\ipb$).The electroweak sector  
chosen is consistent with a LMNDM scenario. The updated results 
corresponding to $\lum$ = 1 $\ifb$ data are presented in Table 6.
     
Finally we look into the $b \tau j$$\etslash$ signal in the mSUGRA 
model (Section 3, Figure 2). We also revisit the $b l j$$\etslash$ 
signal in case 
it gives a better reach in some parameter space. So far as the $\lum$
= 35 $\ipb$ data is concerned both the above signals and the LMNDM 
scenario cannot be strictly ruled out due to the uncertainties in
the data with low statistics ( the ATLAS constraints are reproduced 
in Fig. 2 for ready reference). However, the more recent data for
$\lum = 1 \ifb$ strongly disfavors both the signals and the LMNDM
scenario (see the recent CMS constraints superimposed on  Fig 2).

The latest LHC data also disfavors a large parameter space in mSUGRA with
low mass neutralinos in mSUGRA sensitive to the 
direct DM search by the XENON100 experiment. Alternative models for
LMNDM may, therefore, call for more attention. 

In this paper we have not considered constraints from flavour physics 
like the flavour violating decays of the B-hadrons. Strictly speaking these constraints are 
sensitive to the additional assumption that the quark and the squark mass matrices are 
aligned in the flavour space so that the same matrix as the CKM matrix 
also operate in the squark sector. This assumption of minimum flavour violation
fails even if there are small off-diagonal elements of the
squark mass matrix at the GUT scale. On the other hand such small element does not
affect processes like neutralino annihilation and squark-gluino 
production and decay. In fact, it has been explicitly shown that such small mixings at the GUT scale
can significantly weaken the constraints from flavour physics \cite{okumura}. 
Moreover a comprehensive analysis of all possible constraints on a SUSY model 
is beyond the scope of this paper.

We conclude that any model whose strong sector is not 
sensitive to the current LHC data, may be consistent with a  
LMNDM scenario satisfying the WMAP constraints thanks to the properties 
of the 
EW sector, provided model dependent correlations among the 
parameters of the strong and 
the electroweak sectors, as in mSUGRA, are given up. Thus there is no 
serious conflict between the  LMNDM scenario and the LHC data.  


\end{document}